\documentclass[a4paper,11pt]{article}
\pdfoutput=1 
\usepackage{jheppub} 
                     
\usepackage{booktabs,makecell, bm}
\usepackage[font=small]{caption}
\usepackage{framed}
\usepackage{hyperref}
\usepackage[export]{adjustbox}
\usepackage{amsmath}
\usepackage{lineno}
\usepackage{ytableau}
\usepackage{subfig}
\usepackage{url}
\usepackage[numbers]{natbib}
\usepackage[colorinlistoftodos]{todonotes}

\usepackage{physics}
\usepackage{subcaption}

\title{Equilibrium Partition Function of Non-Relativistic CFTs in Harmonic Trap}

\author{Eunwoo Lee}

\affiliation{Department of Theoretical Physics, \\ Tata Institute of Fundamental Research, Homi Bhabha Rd, Mumbai 400005, India}
\emailAdd{eunwoo.lee@tifr.res.in}

\abstract{We investigate the equilibrium partition function of non-relativistic conformal field theories in harmonic quantization. We first analyze the hydrodynamic regime and show that, at leading order, the partition function exhibits a universal structure determined by the equation of state: the logarithm of the partition function develops simple poles in $\omega^2-\Omega_a^2$, where $\omega$ is the harmonic trapping frequency and $\Omega_a$ are angular velocities acting as chemical potentials for angular momentum. The corresponding residue is determined by a single-variable function of $\mu/T$, with $\mu$ the particle-number chemical potential and $T$ the temperature.

We then study the large-angular-momentum limit where $\mu(\omega-\Omega_a)\ll 1$. In this regime centrifugal effects nearly cancel the trapping potential, and the logarithm of the partition function again exhibits simple poles in $\omega^2-\Omega_a^2$, but with a less universal residue depending separately on $\mu/T$ and $\omega/T$.
As explicit examples we analyze superfluid systems realizable in cold-atom experiments, in particular fermions at unitarity confined in a harmonic trap.}

\makeatletter
\gdef\@fpheader{}
\makeatother

\begin{document} 

\maketitle



\section{Introduction}
The study of partition functions in CFTs has long been a cornerstone for understanding the distribution of operator spectra across various energy scales. It is now a well established fact that the high-temperature behavior of the partition function of a CFT on $S^{d} \times S^1$ follows a universal form where the logarithm of the partition function $\ln Z$ scales with temperature as $T^{d}$, a result primarily driven by dimensional analysis and extensivity.

A system carrying angular momentum is characterized by nonzero chemical potentials $\Omega_i$, which correspond to the angular velocities of the boundary theory. If the system admits a derivative expansion in temperature so that a fluid description applies, the equilibrium partition function of the CFT takes a universal form \cite{Bhattacharyya:2007vs} \footnote{Here we have set the radius of the sphere to unity for simplicity. For more general treatments of the equilibrium partition function using hydrodynamic and effective field theory approaches, see \cite{Banerjee:2012iz,Jensen:2012jh,Benjamin:2023qsc}.}
\begin{align}\label{eq1}
\log Z \sim \frac{T^{d}}{\prod_a (1-\Omega_a)}.
\end{align}

Recent developments in \cite{Anand:2025mfh} instead probe a different region of parameter space, namely the limit in which the angular velocities $\Omega_a$ approach their physical bound, $\Omega_a \to 1$.
In this regime, the system does not necessarily lie within the domain of validity of fluid dynamics.
The authors argue that in this limit, the partition function exhibits a ``semi-universal'' structure, where $\ln Z$ develops a simple pole of the form $(1 - \Omega_a)^{-1}$ for each angular velocity as in \eqref{eq1}. However, the residue of these singularities remains theory-dependent function of $T$ and thus less universal than in \eqref{eq1}, representing the specific dynamical information of the underlying field theory.

Beyond the canonical ensemble, this semi-universality translates into significant constraints on the microcanonical entropy in a specific large spin limit. When $r$ angular momenta $J_1, \dots, J_r$ are scaled to infinity, the entropy $S$ and the twist $\tau\equiv E-\sum_{a=1}^{r} J_a$ also scale to infinity but at a specific relative rate such that the quantities $S / (J_1 \dots J_r)^{1/(r+1)}$ and $\tau / (J_1 \dots J_r)^{1/(r+1)}$ remain fixed. This scaling regime allows for a detailed analysis of phase structures in holographic theories, such as $\mathcal{N}=4$ supersymmetric Yang-Mills theory, where the system transitions between black hole, hairy black hole called grey galaxy \cite{Kim:2023sig,Bajaj:2024utv}, and thermal gas phases \cite{Anand:2025mfh} within this charge regime. These results show that, even in the absence of complete universality in the residues, the functional form of the entropy in terms of the appropriately scaled charges remains strongly constrained by the underlying conformal symmetry.

The universality and semi-universality observed in relativistic CFTs naturally raises the question of whether similar structures exist in theories with non-relativistic conformal symmetry. In this paper we extend that analysis to non-relativistic conformal field theories (NRCFTs) governed by the Schr\"odinger symmetry group. Unlike their relativistic counterparts, NRCFTs possess a conserved particle-number (or mass) operator \(M\) that commutes with the Hamiltonian and other generators. Consequently one must introduce an additional chemical potential \(\mu\) conjugate to the particle number.

We focus primarily on two regimes: (i) the fluid-dynamics (hydro-dynamics) regime and (ii) the large-angular-momentum regime which are counterparts of \cite{Bhattacharyya:2007vs} and \cite{Anand:2025mfh}, respectively. These regimes are not mutually exclusive.

By considering the equilibrium partition function of these theories on a spatial manifold, we investigate whether the approach of angular velocities to their extremal values produces a similar pole structure in the thermodynamic potentials. In the fluid-dynamics regime the partition function takes the form
\begin{align}
\log Z=\beta\int d^d x\,P=
\frac{\mu^d F(\mu/T)}{\prod_{a=1}^r(\omega^2-\Omega_a^2)}\times \begin{cases}
\frac{1}{\omega}, & d=2r+1,\\
1, & d=2r,
\end{cases}
\end{align}
where \(d\) is the spatial dimension and \(F(\mu/T)\) is a dimensionless function.

In the second regime, where we focus on the large angular-momentum limit \(\Omega_a\to\omega\), the equilibrium partition function behaves as
\begin{align}
\log Z \approx \frac{\mu^dF(\mu/T,\omega/T)}{\prod_{a=1}^{r}(\omega^2-\Omega_a^2)}\times \begin{cases}
\frac{1}{\omega}, & d=2r+1,\\
1, & d=2r.
\end{cases}
\end{align}
As in the relativistic case, the residue function \(F\) becomes less universal in this large-angular-momentum limit, encoding more theory-specific dynamical information.
In subsubsection \ref{thermo qu}, we perform a Legendre transform of the partition function to obtain the semi-universal ``extensive" form of the microcanonical entropy:
\begin{align}
S(\tau, Q, J_a) \approx \left( \prod_{a=1}^r J_a \right)^{\frac{1}{r+1}} s_{\text{int}}\left( \frac{\tau}{\left( \prod_{a=1}^r J_a \right)^{\frac{1}{r+1}}}, \frac{Q}{\left( \prod_{a=1}^r J_a \right)^{\frac{1}{r+1}}} \right),
\end{align}
where $\tau$ is the non-relativistic analogue of the twist, defined by $\tau:=E-\omega \sum_a J_a$. The factor $\left( \prod_{a=1}^r J_a \right)^{\frac{1}{r+1}}$ plays the role of an effective spatial volume, while $s_{\text{int}}$ represents a purely dynamical, theory-dependent entropy density.

The paper is organized as follows.
In Section \ref{review}, we briefly review NRCFTs and the Schrödinger algebra.
In Section \ref{fluid}, we derive the equilibrium thermal partition function of NRCFT fluids at large particle number using a derivative expansion.
Section \ref{semiun} shows that the semi-universality of the large-spin limit is a general feature of non-relativistic conformal systems, and identifies new universal scaling behaviors of the entropy in terms of the mass and angular momenta.
Sections \ref{ex1} and \ref{ex2} provide supporting evidence for Section \ref{semiun}: we analyze a free NRCFT and a superfluid NRCFT, respectively, and verify the predicted scaling behavior.
We conclude and discuss some possible future directions in Section \ref{dis}.

\section{Review on Schrödinger Algebra}\label{review}

In this Section, we briefly review Schrödinger Algebra. The Schrödinger algebra is the non-relativistic analog of the conformal algebra. It arises as the symmetry algebra of the free Schrödinger equation and consists of the Galilean algebra plus additional generators for scaling (dilatations), special conformal transformations, and a central extension (mass or particle number).

\subsection*{Galilean Algebra}

The Galilean algebra  (without central extension) includes the following generators:
Hamiltonian  \( H \), momentum 
\( P_i \), spatial rotations
\( M_{ij} \),
and Galilean boosts \( K_i \).

Their commutation relations include:
\begin{align}
    [M_{ij}, P_k] &= i(\delta_{jk} P_i - \delta_{ik} P_j), \nonumber\\
    [M_{ij}, K_k] &= i(\delta_{jk} K_i - \delta_{ik} K_j), \nonumber\\
    [K_i, H] &= i P_i, \nonumber\\
    [K_i, P_j] &= 0 \quad \text{(in the unextended case)}.
\end{align}

To be consistent with quantum mechanics, we allow a central extension:
\begin{align}
    [K_i, P_j] = i \delta_{ij} N,
\end{align}
where \( N \) is a central element, commuting with all other generators. This central extension is necessary to construct non-trivial unitary representations of the Galilean group.

\subsection*{Schrödinger Extensions (Scaling and Conformal-Like Symmetries)}
We add the following generators to extend the algebra: \( D \) dilatation (scale transformations) and \( C \) special conformal transformation.
These form an \( \mathfrak{so}(2,1) \) subalgebra along with \( H \):
\begin{align}
    [D, H] &= 2i H, \nonumber\\
    [D, C] &= -2i C, \\
    [H, C] &= i D.\nonumber
\end{align}
Their commutators with other generators are:
\begin{align}
    [D, P_i] &= i P_i, \nonumber\\
    [D, K_i] &= -i K_i, \\
    [C, P_i] &= i K_i\nonumber.
\end{align}
This structure reflects the anisotropic scaling symmetry with dynamical exponent \( z = 2 \).

\subsection*{Harmonic Quantization and Nonrelativistic Holography}

In relativistic conformal field theories, local operators correspond to states via radial quantization. In non-relativistic theories with Schrödinger symmetry, a similar correspondence exists but is based on evolution in a harmonic potential, often referred to as harmonic quantization.

In a Schrödinger-invariant theory one may introduce a modified Hamiltonian by taking a linear combination of the time-translation generator and the special conformal generator,
\begin{align}
    H_{\text{osc}} = H + \omega^2 C,
\end{align}
where \( \omega \) is a fixed frequency scale. 
The special conformal generator is given by the second spatial moment of the particle-number density,
\begin{align}
    C=\frac{1}{2}\int d^dx\, r^2 \rho(x),
\end{align}
with \( \rho(x) \) the number (mass) density.  Consequently,
\begin{align}
    H_{\text{osc}}
    = H + \frac{\omega^2}{2}\int d^dx\, r^2 \rho(x),
\end{align}
which is precisely the Hamiltonian of the same non-relativistic system coupled to an external harmonic potential \(V(r)=\tfrac12 \omega^2 r^2\).  Thus \(H_{\text{osc}}\) generates time evolution in the harmonically trapped theory.

This can be obtained from the following unitarity transformation
\begin{align}
    H_{\text{osc}}=e^{-i\frac{\pi}{4}\left(\frac{H}{\omega}-\omega C\right)}(\omega D)e^{i\frac{\pi}{4}\left(\frac{H}{\omega}-\omega C\right)}=H+\omega^2C.
\end{align}
The generator \( H_{\text{osc}} \) plays a role analogous to the dilatation operator in radial quantization. The ground state of \( H_{\text{osc}} \) corresponds to the vacuum state, and excited states correspond to insertions of local operators at the origin.

In addition, there exists a state-operator correspondence \cite{Nishida:2007pj}.
To a local operator \( \mathcal{O}(t = 0, \vec{x} = 0) \), one associates a state via
\begin{align*}
    |\mathcal{O}\rangle = \lim_{t \to 0^{-}} \mathcal{O}(t, \vec{x} = 0) |0\rangle.
\end{align*}
These states evolve under \( H_{\text{osc}} \), and their energy eigenvalues correspond to the scaling dimensions \( \Delta \) of the associated operators:
\begin{align*}
    H_{\text{osc}} |\mathcal{O}\rangle = \Delta |\mathcal{O}\rangle.
\end{align*}
The generators \( H, D, C \) form an \( \mathfrak{so}(2,1) \) subalgebra, allowing the organization of operators into representations with highest weight structure. 

The holographic description of a nonrelativistic conformal field theory (NRCFT) is naturally formulated in a spacetime whose isometry group is the Schrödinger group.
Early developments of nonrelativistic holography were initiated in \cite{Son:2008ye,Balasubramanian:2008dm}, and subsequently extended to finite-temperature states described by black hole geometries with Schrödinger asymptotics \cite{Herzog:2008wg,Maldacena:2008wh,Adams:2008wt}.
Such geometries possess a null Killing direction, and the conserved momentum along this direction is identified with the particle number of the dual theory. Geometrically, Schrödinger spacetimes belong to the Bargmann class: a null reduction along the Killing direction produces Galilean-invariant dynamics on the boundary.

In global Schrödinger coordinates the metric may be written as \cite{Blau:2009gd}
\begin{align}
ds^2
= -\frac{dT^2}{R^4}
+ \frac{1}{R^2}\Bigl(
-2\,dT\,dV
- \omega^2 (R^2 + X^2)\, dT^2
+ dR^2 + dX^2
\Bigr),
\end{align}
where the conformal boundary is located at \(R=0\). The parameter \(\omega\) corresponds to a harmonic trapping potential in the dual field theory and implements the state--operator correspondence of the NRCFT. Translations in the time coordinate \(T\) generate evolution with respect to the trapped Hamiltonian \(H_{\rm osc}\).

\section{Rotating non-relativistic fluid}\label{fluid}

In this section we study non-relativistic fluids and construct the equilibrium thermal partition function of a rotating non-relativistic fluid.
Following \cite{Bhattacharyya:2007vs} for relativistic CFTs, we construct nonlinear rotating solutions of non-relativistic fluid dynamics, which in a holographic setting would correspond to rotating black hole configurations in the bulk.

Within holography, the hydrodynamics of an NRCFT is expected to be described by long-wavelength perturbations of an appropriate bulk black hole geometry \cite{Rangamani:2008gi}. While this suggests the existence of a gravitational dual—though none is currently known to the author—our analysis below relies only on general hydrodynamic principles and does not require specifying a particular bulk solution. \footnote{The black hole geometries with nonrelativistic asymptotics constructed in \cite{Herzog:2008wg,Maldacena:2008wh,Adams:2008wt} need not directly realize the fluid regime considered here. In particular, the thermodynamic ensemble relevant for our fluid requires a positive particle-number chemical potential $\mu$, whereas the standard parametrizations of those solutions typically correspond to a different range of chemical potentials.}

Consider a non-relativistic Galilean-invariant fluid in $d$ spatial
dimensions, described by the local fields
\[
\rho(t,\mathbf{x}), \qquad v^i(t,\mathbf{x}), \qquad T(t,\mathbf{x}), \qquad s(t,\mathbf{x}) .
\]
where $\rho$ is the mass (number) density, $v^i$ is the fluid velocity field, $T$ is the local temperature, and $s$ is the entropy density.
The system is placed in a harmonic potential
\begin{equation}
V(\mathbf{x}) = \frac{1}{2}\,\omega^2 r^2 ,
\end{equation}
and we work in flat space.

The conservation laws take the form
\begin{align}
\partial_t \rho + \partial_i(\rho v^i) &= 0, \label{eq:mass} \\
\partial_t(\rho v^i) + \partial_j \Pi^{ij} &= - \rho\,\partial^i V, \label{eq:mom} \\
\partial_t \varepsilon + \partial_i j_\varepsilon^i
&= - v^i \rho\,\partial_i V . \label{eq:energy}
\end{align}
where $\Pi^{ij}$ is the stress tensor, $\varepsilon$ is energy density, and $j_\varepsilon^i$ is spatial energy current. $\Pi^{ij}$ is written to first order in derivatives as
\begin{equation}
\Pi^{ij}
=
\rho v^i v^j
+ P\,\delta^{ij}
- \eta\,\sigma^{ij}
- \zeta\,\delta^{ij}\,\partial_k v^k.
\end{equation}
$\sigma^{ij}$ is shear tensor is
\begin{equation}
\sigma^{ij}
=
\partial^i v^j + \partial^j v^i
- \delta^{ij}\,\partial_k v^k .
\end{equation}
The energy current is
\begin{equation}
j_\varepsilon^i
=
(\varepsilon + P)v^i
- \kappa\,\partial^i T .
\end{equation}

The entropy current is written as
\begin{equation}
j_s^i = s\,v^i - \frac{\kappa}{T}\,\partial^i T ,
\end{equation}
whose divergence is non-negative
\begin{equation}
\partial_t s + \partial_i j_s^i
=
\frac{\eta}{2T}\sigma^{ij}\sigma_{ij}
+
\frac{\zeta}{T}(\partial_i v^i)^2
+
\frac{\kappa}{T^2}(\partial_i T)^2
\;\ge\;0 .
\end{equation}
Exact equilibrium (zero entropy production) therefore requires
\begin{equation}\label{eqcon}
\sigma^{ij}=0,
\qquad
\partial_i v^i=0,
\qquad
\partial_i T=0 .
\end{equation}
Under these conditions all dissipative and diffusive contributions vanish,
and the fluid is non-dissipative. Imposing the above constraints, the stress tensor, momentum density,
energy density, and energy current reduce to
\begin{equation}
\begin{aligned}
\Pi^{ij} &= \rho\, v^i v^j + P\,\delta^{ij}, \\[4pt]
\pi^i &= \rho\, v^i, \\[4pt]
\varepsilon
&=
\frac{1}{2}\rho\, v^2
+ \varepsilon_{\mathrm{int}}(\rho)
+ \frac{1}{2}\omega^2 r^2\,\rho, \\[4pt]
j_\varepsilon^{\,i} &= (\varepsilon + P)\, v^i .
\end{aligned}
\end{equation}

\subsection{Shearless and divergence-free stationary flows}

We consider a velocity field $v^i(x)$ in $d$-dimensional flat space satisfying \eqref{eqcon}:
\begin{align}\label{eq:divfree}
\partial_i v^i &= 0, \nonumber\\
\sigma^{ij}&\equiv
\partial^i v^j + \partial^j v^i
- \frac{2}{d}\delta^{ij}\,\partial_k v^k
=0.
\end{align}
Since the flow is divergence-free, the shear-free condition reduces to
\begin{align}
\partial^i v^j + \partial^j v^i = 0 .
\end{align}
Thus $v^i$ is a Killing vector of flat space.

The general smooth solution on $\mathbb{R}^d$ is therefore linear in the coordinates,
\begin{align}
v^i(x) = \Omega^{ij} x_j + b^i ,
\qquad
\Omega^{ij}=-\Omega^{ji},
\end{align}
where $b^i$ is a constant vector generating translations and $\Omega^{ij}$ is a constant antisymmetric matrix generating rotations. The translational part must be set to zero to satisfy the stationarity condition for a fluid in a harmonic trap.

Any antisymmetric matrix $\Omega^{ij}$ can be brought, by an orthogonal transformation, to a block-diagonal canonical form. Writing
\begin{align}
r=\left\lfloor \frac d2 \right\rfloor ,
\end{align}
the velocity field can be expressed in suitable coordinates as independent rigid rotations in $r$ mutually orthogonal planes,
\begin{align}
\begin{aligned}
v^1 &= -\Omega_1 x^2, & v^2 &= \Omega_1 x^1,\\
v^3 &= -\Omega_2 x^4, & v^4 &= \Omega_2 x^3,\\
&\vdots
\end{aligned}
\end{align}
with angular velocities $\Omega_a$ $(a=1,\dots,r)$. For odd $d$, there is one additional direction with vanishing velocity.

This shows that, up to coordinate rotations, rigid rotations in orthogonal planes with constant angular velocities exhaust all stationary, divergence-free, and shearless velocity fields in $d$ dimensions.

\subsection{Stationary solution with unequal angular velocities}

We  consider the stationary Euler equation
\begin{align}
\rho\, v^j\partial_j v^i + \partial^i P = -\rho\,\partial^i V ,
\label{eq:euler_general}
\end{align}
with an isotropic harmonic trap
\begin{align}
V=\frac12 \omega^2 r^2 .
\end{align}
The convective term evaluates to
\begin{align}
v^j\partial_j v^i
=
\Omega^{ik}\Omega_{kj}x^j .
\end{align}
In the above canonical basis this becomes, in each rotation plane,
\begin{align}
v^j\partial_j v^i = -\Omega_a^2 x^i
\qquad
(i=2a-1,\,2a),
\end{align}
and vanishes along the non-rotating direction if $d$ is odd.

The Euler equation \eqref{eq:euler_general} therefore reduces to
\begin{align}
\partial^i P
=
\rho\,\left(\Omega_a^2-\omega^2\right)x^i ,
\qquad
i=2a-1,\,2a,
\label{eq:anisotropic_gradP}
\end{align}
with $\partial^i P=-\rho\,\omega^2 x^i$ along any non-rotating direction.

\subsubsection{Density profile}
Schr\"odinger conformal invariance together with extensivity fixes the equation of state to take the form
\begin{align}
P=\rho^{1+\frac{2}{d}}\, g\!\left(\frac{\rho^{2/d}}{T}\right),
\label{eq:schrod_eos}
\end{align}
where $g(\rho^{2/d}/T)$ is a theory-dependent function.
\footnote{Strictly speaking, the system is placed in a harmonic trapping potential, so translation invariance is broken and global extensivity does not hold. Instead, extensivity applies only locally, in the sense of the local density approximation. This situation is analogous to that of a relativistic conformal fluid on a sphere, where the fluid in principle couples to the background curvature, but curvature effects are negligible at leading order in the hydrodynamic regime. This approximation is valid when the microscopic mean free path length is much smaller than the characteristic length scale of the harmonic trap. We will discuss the corresponding regime of validity at the end of this section.}
Using the Euler equation \eqref{eq:anisotropic_gradP}, we obtain
\begin{align}\label{soleq}
\frac{\partial P}{\partial x^i} = \frac{\partial P}{\partial\rho}\frac{d \rho}{d x^i}
=
\rho\,(\Omega_a^2-\omega^2)\,x^i.
\end{align}
Dividing the equation by $\rho$ and integrating in $x$, we observe that the charge density
$\rho$ is given as a function of $\mu-\frac{1}{2}Q(x)$ and $T$ as follows 
\begin{align}\label{rhorho}
    \rho=\mathcal{G}(\mu-\frac{1}{2}Q(x),T)=(\mu-\frac{1}{2}Q(x))^{d/2}\tilde{\mathcal{G}}\left(\frac{\mu-\frac{1}{2}Q(x)}{T}\right),
\end{align}
where $\mathcal{G}$ is a function that is given by solving \eqref{soleq}, and $Q(x)$ given by
\begin{align}
Q(x)
=
\sum_{a=1}^{r}(\omega^2-\Omega_a^2)(x_{2a-1}^2+x_{2a}^2)
+
\begin{cases}
\omega^2 x_{2r+1}^2, & d=2r+1,\\
0, & d=2r,
\end{cases}
\end{align}
The fluid occupies an ellipsoidal region defined by $\rho(x)\ge 0$. The parameter $\mu$ is fixed such that $\rho(x)$ vanishes when $\mu-\frac{1}{2}Q(x)\le 0$. One may interpret $\mu-\frac{1}{2}Q(x)$ as a local chemical potential $\mu_{\rm eff}$. In fact, \eqref{soleq} is equivalent to $\frac{\partial P}{\partial\mu_{\rm eff}}=\rho(x)$, which is a local thermodynamic relation. 

\subsubsection{Pressure profile and integrated pressure}

The pressure profile follows directly from \eqref{eq:schrod_eos} and \eqref{rhorho}:
\begin{align}
P(x)
&=\left(\mu-\frac{1}{2}Q(x)\right)^{1+\frac{d}{2}}
\,f\!\left(\frac{\mu-\frac{1}{2}Q(x)}{T}\right),
\end{align}
valid for $2\mu\ge Q(x)$.  For $2\mu<Q(x)$ the density is exponentially suppressed, so one may effectively take $P(x)\approx 0$ (equivalently $\rho(x)\approx 0$).
Since $\int d^dx P$ is equivalent to $\frac{1}{\beta}\log Z$, thermodynamic quantities are given by the integrated pressure.
To compute this, we rescale coordinates
\begin{align}
y_{2a-1,2a}
=
\sqrt{\omega^2-\Omega_a^2}\,x_{2a-1,2a},
\qquad
y_{2r+1}=\omega x_{2r+1}\ \ (d\ \text{odd}),
\end{align}
which maps the integration domain from an ellipsoid to a ball. The result is
\begin{align}\label{logz}
\log Z=\beta\int d^d x\,P=F(\mu/T)\times 
\frac{\mu^d}{\prod_{a=1}^r(\omega^2-\Omega_a^2)}\times \begin{cases}
\frac{1}{\omega}, & d=2r+1,\\
1, & d=2r,
\end{cases}
\end{align}
where $F(\mu/T)$ is a dimensionless function of $\mu/T$ obtained after integrating $P$.

Conserved charges are given as follows. For notational convenience we write for $d=2r$.
\begin{align}\label{cons charge}
J_a &= T\frac{\partial \log Z}{\partial \Omega_a}= \frac{2T\Omega_a}{\omega^2-\Omega_a^2}
\frac{\mu^dF(\mu/T)}{\prod_{b=1}^{r}(\omega^2-\Omega_b^2)},\nonumber\\
Q &= T\frac{\partial \log Z}{\partial\mu},\nonumber\\
S &= \log Z + T\frac{\partial \log Z}{\partial T}=\frac{\mu^d(F(\mu/T)+T\partial_T F(\mu/T))}{\prod_{b=1}^{r}(\omega^2-\Omega_b^2)},\nonumber\\
E &= T^2\frac{\partial \log Z}{\partial T}
+\mu Q
+\sum_{a=1}^{r}\Omega_aJ_a=dT\mu^{d}
\frac{F(\mu/T)}{\prod_{a=1}^r(\omega^2-\Omega_a^2)}+\sum_{a=1}^{r}\Omega_aJ_a
\end{align}
One can compute all thermodynamic quantities once the function $F(\mu/T)$, which depends on a single parameter, is specified.

\subsubsection*{Validity Regime}
Let us discuss when the fluid dynamics is valid. In general, fluid dynamics is valid when the mean free path $l_{\rm mfp}$ is much smaller than the characteristic macroscopic length scale $L$ of the system. Namely, the expansion parameter controlling hydrodynamics is the Knudsen number
\begin{equation}
\mathrm{Kn} \sim \frac{\ell_{\rm mfp}}{L}
\end{equation}
which we require it to be small for the fluid description to be valid.

Momentum transport is governed by the momentum diffusion constant
\begin{equation}
D_\pi \sim \frac{\eta}{\rho},
\qquad \text{(units: length$^{2}$/time)}.
\end{equation}
The associated microscopic length scale (mean free path) is obtained by dividing by a typical excitation velocity $v$,
\begin{equation}
\ell_{\rm mfp} \sim \frac{D_\pi}{v}
\sim \frac{\eta}{\rho\, v}.
\end{equation}
where Knudsen number is given by ${\rm Kn}=\frac{s}{\rho}\frac{1}{vL}$.
(In relativistic units, where $v\sim 1$, this reduces to the familiar estimate $\ell_{\rm mfp}\sim \eta/\varepsilon$.)

If we pay our attention to a strongly coupled holographic theory, the ratio $\eta/s$ is parametrically constant, so that the mean free path $l_{\rm mfp}$ is given as
\begin{equation}
\ell_{\rm mfp} \sim \frac{\eta}{\rho\, v}
\sim \text{const}\times \frac{s}{\rho\, v}.
\end{equation}
From dimensional analysis, $v\sim\sqrt{\mu}$ and the system size $L\sim \frac{v}{\omega} \sim \frac{\sqrt{\mu}}{\omega}$. Therefore, the Knudsen number, which must be small for hydrodynamics to apply, is
\begin{align}\label{knud}
\frac{\ell_{\rm mfp}}{L}
\sim
\frac{s}{\rho}\cdot \frac{\omega}{\mu}\ll1.
\end{align}

One can compute the ratio $s/\rho\sim S/Q$ from \eqref{cons charge}. For a fluid with $\mu\sim T$, $Q$ and $S$ are of the same order for a generic function $F(\mu/T)$, so one expects $\omega/\mu$ to be small for hydrodynamics to be valid.

On the other hand, if we are describing a near-extremal fluid with a finite entropy (holographically dual to an extremal black hole), the pressure should be analytic in $T$ at low temperature, so $F(\mu/T)$ in \eqref{logz} can be written as
\begin{equation}
F(\mu/T)=A \frac{\mu}{T}+ B + O\left(\frac{T}{\mu}\right).
\end{equation}
It follows that
\begin{equation}
Q\propto A\mu^{d} ,
\qquad
S \propto  B\,\mu^{d},
\end{equation}
and therefore
\begin{equation}
\frac{s}{\rho}
\propto
\frac{B}{A},
\end{equation}
such that $S/Q$ is constant.
In this case, $\omega/\mu$ should be small for the Knudsen number to be small.

\section{Semi-universality at large angular momentum}\label{semiun}

In this section we study the thermal equilibrium partition function of the system in the regime of large angular momentum. Our main result is that, in this limit, the free energy takes the following “semi-universal” form:
\begin{align}\label{logz2} 
\log Z=F(\mu/T,\omega/T)\times \frac{\mu^d}{\prod_{a=1}^r(\omega^2-\Omega_a^2)}\times \begin{cases} \frac{1}{\omega}, & d=2r+1,\\ 1, & d=2r, \end{cases} 
\end{align}
where, in contrast to \eqref{logz}, the function $F(\mu/T,\omega/T)$ does not depend on a single parameter, but depends separately on two dimensionless parameters $\mu/T$ and $\omega/T$. As we will show, in the canonical ensemble, large angular momentum is obtained by taking the limits
$\beta(\omega-\Omega)\to 0$ and $\mu(\omega-\Omega)\to 0$. This regime does not necessarily lie within the domain of validity of fluid dynamics.

As explained in the previous section, the fluid description applies when the Knudsen number is small. For example, in holographic theories with $\mu\sim T$, this requires $\omega\ll\mu$, as follows from \eqref{knud}, and the function $F(\mu/T,\omega/T)$ reduces to $F(\mu/T)$.

Beyond the fluid-dynamical regime, the standard derivative expansion is no longer strictly valid, making the analysis more difficult. Nevertheless, following the strategy used in the relativistic case in \cite{Anand:2025mfh}, we formally carry out the expansion as if the system were still within the fluid-dynamical regime. 
We then verify that, at each order in the expansion, the most divergent contributions reproduce the universal divergence structure shown in \eqref{logz2}.

Although the series of the derivative expansion may not be convergent, the fact that every term in the infinite sequence exhibits the same scaling behavior in the $\Omega\to\omega$ limit suggests that these contributions can be resummed into a theory-dependent function of the thermodynamic parameters, $F(\mu/T,\omega/T)$, which encodes the underlying dynamics of the CFT. This uniform scaling across all orders provides a justification for the validity of the resulting expression beyond the strict fluid-dynamical regime.

\subsection{Derivative expansion of the equilibrium partition function for Schrödinger CFTs}

In complete analogy with relativistic conformal field theories, we construct the
equilibrium partition function of a Schrödinger-invariant field theory as a functional
of background geometric data and perform a derivative expansion.
The appropriate background geometry for Schrödinger CFTs is Newton-Cartan geometry.

First, we consider so-called the Bargmann background geometry, which is described by the \((d+2)\)-dimensional Lorentzian metric
\begin{align}\label{Bargmann}
ds^2 = 2n_{\mu}dx^{\mu}\bigl(dx^-+m_{\mu}dx^{\mu}\bigr) + h_{\mu\nu}dx^{\mu}dx^{\nu},
\end{align}
which possesses a null isometry generated by \(\partial_-\).

Upon null reduction, the metric becomes Newton-Cartan and momentum along the \(x^-\) direction becomes a central charge and is identified with the particle number (or mass) in the non-relativistic CFT. In this decomposition,
\(n_\mu\) is the thermal clock one-form,
\(m_\mu\) is the particle-number gauge field, whose temporal component plays the role of the chemical potential, and
\(h_{\mu\nu}\) is the spatial metric, satisfying
\[
h^{\mu\nu}n_\nu = 0 .
\]
This structure is the Schrödinger analogue of the Kaluza–Klein metric appearing in the relativistic case.

The thermal partition function is obtained by Wick-rotating to Euclidean time, \(t\to -i\tau\), leading to
\begin{align}\label{part}
    Z = \mathrm{Tr} e^{-\beta(H - \mu N)}
\end{align}
with the Euclidean time coordinate periodically identified as \(\tau\sim\tau+\beta\).

If one further introduces angular momenta and their associated chemical potentials, the grand-canonical partition function takes the form
\begin{align}
Z=\mathrm{Tr} e^{-\beta\left(H-\mu N-\sum_{a}\Omega_a J_a\right)} ,
\end{align}
where \(\Omega_a\) are the angular velocities conjugate to the conserved angular momenta \(J_a\). In the Euclidean formulation, these chemical potentials are implemented by twisted boundary conditions on the angular coordinates, $\theta_a\sim\theta_a-i\Omega_a\beta$.

Here we decompose $\mathbb{R}^d$ into mutually orthogonal rotation planes by writing
\(d=2r+\epsilon,~\epsilon\in\{0,1\}.\)
For each plane we introduce polar coordinates
\begin{align}
x_{2a-1}=r_a\cos\theta_a,\qquad
x_{2a}=r_a\sin\theta_a,\qquad a=1,\dots,r,
\end{align}
so that $x_{2a-1}^2+x_{2a}^2=r_a^2$. 
If $d$ is odd ($\epsilon=1$) we denote the remaining coordinate by
$z\equiv x_{2r+1}$.
In these coordinates the flat metric becomes
\begin{align}
\sum_{i=1}^{d} dx_i^2
=
\sum_{a=1}^{r}\left(dr_a^2+r_a^2\,d\theta_a^2\right)
+\epsilon\,dz^2.
\end{align}

For stationary backgrounds one may choose coordinates such that
\begin{equation}\label{coord}
\begin{aligned}
n_\mu dx^\mu &= e^{\sigma(x)}\bigl(d\tau + a_i(x)\,dx^i\bigr), \\
h^{ij} &= g^{ij}(x), \\
m_\mu dx^\mu &= \mu_{\rm eff}(x)\,d\tau + \tilde m_i(x)\,dx^i ,
\end{aligned}
\end{equation}
For simplicity, let $d=2r$ is even and assume equal angular momenta $\Omega_a=\Omega$. Then, define $\phi_a$ such that 
\begin{align}
    \phi_a\equiv\theta_a+i\Omega \tau.
\end{align}
A nice feature of this coordinate transformation is that, since the $\phi_a$ are periodic, we are effectively describing a partition function without $\Omega$, as in \eqref{part}.
The the background metric in terms of $\tau,r_a,\phi_a$ is written as:
\begin{align}\label{newton metric}
    ds^2&=2n_{\mu}dx^{\mu}(dx^-+m_{\mu}dx^{\mu})+h_{\mu\nu}dx^{\mu}dx^{\nu}\nonumber\\
    &=2(-id\tau)(dx^{-}-i(\mu-\frac{\omega^2r^2}{2})d\tau)+\sum_a(dr_a^2+r_a^2(d\phi_a-i\Omega d\tau)^2)\nonumber\\
    &=2(-id\tau)(dx^{-}-i(\mu-\frac{(\omega^2-\Omega^2)r^2}{2})d\tau+\Omega r^2 \sum_a d\phi_a)+\sum_a(dr_a^2+r_a^2d\phi_a^2)
\end{align}
where $r^2\equiv\sum_{a}r_a^2$ and we get the gauge field for the mass:
\begin{align}
    m_{\mu}dx^{\mu}=(\mu-\frac{(\omega^2-\Omega^2)r^2}{2})dt+\Omega r^2 \sum_{a}d\phi_a
\end{align}
Therefore, 
\begin{align}
    \mu_{\rm eff}(x)=\mu-\frac{(\omega^2-\Omega^2)r^2}{2},\qquad \tilde{m}_\phi= \Omega r^2, \qquad
    h_{\mu\nu}dx^{\mu}dx^{\nu}=\sum_{a}(dr_a^2+r_a^2d\phi_a^2)
\end{align}
with $\sigma, a_i$ being zero.

In general dimensions with different angular velocities $\Omega_a$ $(a=1,\dots,r)$,
$r=\lfloor d/2\rfloor$, the effective chemical potential is given as
\begin{equation}
\mu_{\rm eff}(x)
=
\mu
-
\frac12
\sum_{a=1}^{r}
\left(\omega^2-\Omega_a^2\right)
\left(x_{2a-1}^2+x_{2a}^2\right)
-
\begin{cases}
\frac12 \omega^2 x_{2r+1}^2, & d=2r+1,\\
0, & d=2r .
\end{cases}
\end{equation}
and the spatial metric $h$ is that of a flat $d$-dimensions.

\subsubsection*{Euclidean partition function}

The Euclidean partition function is defined as a path integral on this background,
\begin{equation}
Z[n_\mu,h^{\mu\nu},m_\mu]
=
\int \mathcal D\Phi\;
e^{-S_E[\Phi;n_\mu,h^{\mu\nu},m_\mu]} .
\end{equation}
Stationarity implies that $\log Z$ reduces to a local functional of spatial fields.

One may write the partition function as
\begin{equation}
\log Z
=
\int d^d x\,\sqrt{g}~
\mathcal W ,
\end{equation}
where $d$ is the number of spatial dimensions of the NRCFT.
$\sqrt{g}=\sqrt{h_{ij}}$ (cf. \eqref{coord})

$\mathcal W$ admits a derivative expansion,
\begin{equation}
\mathcal W
=
\mathcal W^{(0)}
+
\mathcal W^{(1)}
+
\mathcal W^{(2)}
+\cdots .
\end{equation}
Here $\mathcal W^{(n)}$ contains terms with $n$ spatial derivatives.

\subsubsection*{Zeroth order}

The background fields consist of the clock one-form $n_\mu$, the spatial metric
$g_{ij}$, and the mass gauge field $m_\mu$. A term in the equilibrium generating
functional must be invariant under spatial diffeomorphisms, $U(1)_M$ mass gauge
transformations, and anisotropic Weyl transformations. At zeroth order, no
derivatives of background fields are allowed.

The local temperature is defined geometrically by the proper length of the
Euclidean time circle,
\begin{equation}
T(x) = \frac{1}{\beta\, n_0(x)} = \frac{e^{-\sigma(x)}}{\beta},
\end{equation}
where $n_0=e^{\sigma}$ in our parametrization. Under anisotropic Weyl
transformations, $n_\mu\to e^{2\phi}n_\mu$, the temperature transforms with Weyl
weight $-2$.

The background field component $\mu(x)$ appearing in
$m_\mu dx^\mu=\mu(x)\,d\tau+\tilde m_i dx^i$ is not gauge invariant by itself and
therefore cannot appear alone in the generating functional. 
\begin{align}
    m_{\mu}\to m_{\mu}+\partial_{\mu}\Lambda
\end{align}
The gauge-invariant
and diffeomorphism-invariant chemical potential is instead given by the
contraction
\begin{equation}
\nu(x) \equiv \oint_{\tau} dx^{\mu} m_\mu 
\end{equation}
which represents the local energy shift of charged states. This quantity has Weyl weight $0$.

Schrödinger scale invariance in $d$ spatial dimensions fixes the form $P(T,\mu)
=
T^{\frac d2}\,
f\!\left(\nu\right)$ such that 
\begin{equation}\label{zero}
\log Z^{(0)}=\int d^d x\,\sqrt{g}~ T^{\frac{d}{2}}f(\nu)
\end{equation}
is Weyl invariant.

In the system of interest, $\nu=\frac{\mu-(\omega^2-\Omega^2)r^2/2}{T}$.

\subsubsection*{First order}

At first order in derivatives, no scalar invariants consistent with
rotational invariance, time-reversal symmetry, and Schrödinger symmetry exist.
Therefore, $\mathcal W^{(1)} = 0 .$ (Note that we are only considering parity invariant theories in this paper)

\subsubsection*{Second order}

At second order, allowed scalar invariants are
\begin{equation}
\begin{aligned}
(\nabla_i \sigma)^2, \qquad
(\nabla_i \nu)^2, \qquad
f_{ij}f^{ij}, \qquad
\tilde f_{ij}\tilde f^{ij},\qquad
R[g]
\end{aligned}
\end{equation}
where $
f_{ij}=\partial_i a_j-\partial_j a_i,~
\tilde f_{ij}=\partial_i \tilde m_j-\partial_j \tilde m_i,$ and $R[g]$ is a Ricci scalar.
The second-order contribution therefore takes the form
\begin{equation}
\mathcal W^{(2)}
=
\alpha_1(T,\mu)(\nabla\sigma)^2
+
\alpha_2(T,\mu)(\nabla\nu)^2
+
\alpha_3(T,\mu) f_{ij}f^{ij}
+
\alpha_4(T,\mu) \tilde f_{ij}\tilde f^{ij}
+
\alpha_5(T,\mu) R[g].
\end{equation}

We impose invariance of the equilibrium generating functional under
local anisotropic Weyl transformations,
\begin{equation}
\begin{aligned}
\sigma &\to \sigma + 2\phi(x), \\
g_{ij} &\to e^{2\phi(x)} g_{ij}, \\
a_i &\to a_i, \\
m_\mu &\to m_\mu,
\end{aligned}
\end{equation}
requiring that the variation of the partition function vanish up to total
derivatives,
\begin{equation}
\delta_\phi \log Z
=
\int d^d x \sqrt{g}\; \delta_\phi \mathcal W
= 0 .
\end{equation}

The relevant Weyl variations are
\begin{align}
\delta_\phi (\nabla_i \sigma)
&= 2\nabla_i \phi , \\
\delta_\phi (\nabla \sigma)^2
&=
4\,\nabla^i\sigma\nabla_i\phi
+
4(\nabla\phi)^2 , \\
\delta_\phi (\nabla\nu)^2
&=
-2\phi(\nabla\nu)^2 , \\
\delta_\phi (f_{ij}f^{ij})
&=
-4\phi\, f_{ij}f^{ij} ,
\end{align}
with an identical transformation for $\tilde f_{ij}\tilde f^{ij}$.
Only $(\nabla\sigma)^2$ produces inhomogeneous terms involving derivatives of
$\phi$.

To achieve Weyl invariance at second order, it is necessary to include the
spatial Ricci scalar $R[g]$, which is an allowed second-order scalar in
Newton–Cartan geometry. We therefore extend
\begin{equation}
\mathcal W^{(2)}
\supset
\alpha_5(T,\mu)\, R[g] .
\end{equation}
Its Weyl variation in $d$ spatial dimensions is
\begin{equation}
\delta_\phi R
=
-2\phi R
-2(d-1)\nabla^2\phi
-(d-1)(d-2)(\nabla\phi)^2 .
\end{equation}
After integrating by parts, the $\nabla^2\phi$ term generates
$\nabla\sigma\cdot\nabla\phi$ contributions, while the $(\nabla\phi)^2$ term
allows cancellation of the inhomogeneous variation from $(\nabla\sigma)^2$.

Requiring cancellation of all inhomogeneous terms yields
\begin{align}
4\alpha_1
-
(d-1)(d-2)\alpha_5
&= 0 , \\
4\alpha_1
-
2(d-1)\frac{\partial \alpha_5}{\partial \sigma}
&= 0 .
\end{align}
Using
\begin{equation}
\frac{\partial}{\partial\sigma}
=
-2T\frac{\partial}{\partial T} ,
\end{equation}
these conditions are consistent and imply the constraint
\begin{equation}
\alpha_1
=
\frac{(d-1)}{2(d-2)}\,
T\,\partial_T \alpha_5 .
\end{equation}
$d=2$ is special; $\alpha_1$ must be zero.

Homogeneous Weyl scaling further fixes the temperature dependence of the
coefficients:
\begin{equation}
\begin{aligned}
\alpha_1(T,\mu),\;\alpha_2(T,\mu),\;\alpha_5(T,\mu)
&\sim T^{\frac d2-1}~ f(\mu/T), \\
\alpha_3(T,\mu),\;\alpha_4(T,\mu)
&\sim T^{\frac d2-2}~f(\mu/T),
\end{aligned}
\end{equation}
up to arbitrary functions of the Weyl-invariant ratio $\beta\,\mu_{\rm therm}(x)$.

The most general second-order Weyl-invariant contribution to the generating
functional is therefore
\begin{equation}\label{second}
\begin{aligned}
\mathcal W^{(2)} &=
\alpha_1(T,\mu)(\nabla\sigma)^2
+
\alpha_2(T,\mu)(\nabla\nu)^2
\\
&\quad
+
\alpha_3(T,\mu) f_{ij}f^{ij}
+
\alpha_4(T,\mu)\tilde f_{ij}\tilde f^{ij}
+\alpha_5(T,\mu)\, R[g]
,
\end{aligned}
\end{equation}
with $\alpha_1$ constrained as above.

\subsubsection*{Higher order}

One can do the similar analysis for higher orders in derivatives and only even-order terms contribute if the theory is parity invariant.
The building blocks are as follows.
\begin{align}
    \nabla\sigma, \quad\nabla\nu,\quad f_{ij}, \quad\tilde{f}_{ij}, \quad \text{and~}R_{ijkl}.
\end{align}
Each object carries one derivative (with the Riemann tensor effectively counting as two derivatives). At fourth order, one may therefore consider, for example,
$$R_{ij}R^{ij}, (\nabla_kf_{ij})^2, R\tilde{f}_{ij}\tilde{f}^{ij}, (\nabla\sigma)^2f_{ij}f^{ij},$$
and similar contractions.

\subsection{Semi-universality in free energy at large spin}

From now on we focus on the Newton-Cartan metric written in\eqref{newton metric}.
In the local density approximation, the free energy takes the form
\begin{equation}
\log Z
=
\int d^d x\,\sqrt{g}\;
T^{d/2}\, f\!\left(\nu(x)\right)
+\int d^d x\,\sqrt{g}\;
\mathcal{W}^{(2)}
+\cdots ,
\end{equation}
where \(\nu(x)=\beta\mu_{\rm eff}(x)\) and \(\mu_{\rm eff}(x)\) is the local chemical potential.

As in Section \ref{fluid}, we redefine the coordinates as
\begin{align}
    y_{2a-1}^2\equiv (\omega^2-\Omega_a^2)x_{2a-1}^2,\qquad
    y_{2a}^2\equiv (\omega^2-\Omega_a^2)x_{2a}^2.
\end{align}
The zeroth-order contribution to the free energy then becomes
\begin{align}
    \int d^d x\,\sqrt{g}\;\mathcal{W}^{(0)}
    =
    \frac{1}{\prod_{a=1}^{r}(\omega^2-\Omega_a^2)}
    \int_{0}^{\infty} d^d y\, T^{d/2}
    f\!\left(\frac{\mu-\frac{1}{2}r^2}{T}\right)
    =
    \frac{\mu^{d}F_0(\mu/T)}{\prod_{a=1}^{r}(\omega^2-\Omega_a^2)} ,
\end{align}
which is proportional to \(\prod_{a=1}^{r}(\omega^2-\Omega_a^2)^{-1}\).

As a slight digression, we distinguish two cases depending on the sign of the chemical potential at the trap center.
When \(\mu_{\rm eff}(x)<0\), the density is dilute and typically exponentially suppressed, like
\begin{align}
    \rho(x)\sim T^{d/2}\exp\!\left[\frac{\mu_{\rm eff}(x)}{T}\right].
\end{align}
When \(\mu_{\rm eff}(x)>0\), the energy density is an increasing function of the particle density,
\begin{align}
    \mu_{\rm eff}(x)=\frac{\partial \varepsilon}{\partial\rho}>0.
\end{align}
The fluid dynamics regime belongs to the chemical potential at the center $\mu_{\rm eff}(0)=\mu$ positive. In the previous section we have approximated the particle density as vanishing whenever
\(\mu_{\rm eff}(x)\le 0\) (strictly speaking, it decays exponentially outside this region and is therefore negligible). The fluid thus occupies a finite ellipsoidal region defined by
\begin{equation}
\sum_{a=1}^{r}
\left(\omega^2-\Omega_a^2\right)
\left(x_{2a-1}^2+x_{2a}^2\right)
+
\begin{cases}
\omega^2 x_{2r+1}^2, & d=2r+1,\\
0, & d=2r
\end{cases}
\le
2\mu .
\end{equation}

In any case, we now show that the maximal divergence of the second–order terms is of the same order as that of the zeroth–order contribution. Only the second and fourth terms in \eqref{second} contribute, since $f_{ij}=0$ and the spatial part of the Riemann tensor is zero. The second term $(\nabla \nu_i)^2$ scales as $(\omega^2-\Omega^2)/\mu$. Moreover, the coefficient \(\alpha_2\) is suppressed by an additional factor of \(1/\mu\) relative to the zeroth-order contribution,\footnote{Since \(T^{d-1}f(\mu/T)\) may be written as \(\mu^{d-1}\tilde f(\mu/T)\). Here we implicitly assume \(f\sim\tilde f\sim O(1)\), i.e. \(\mu\sim T\). The unbalanced limit \(T\to 0\) is discussed in Section \ref{ex2}. At the end of Section \ref{fluid} we also briefly discuss the holographic fluid at zero temperature.}\label{3} so that the resulting relative scaling is \((\omega^2-\Omega^2)/\mu^2\).

On the other hand, the fourth term satisfies $\tilde{f}_{ij}\tilde{f}^{ij}=\Omega^2\approx \omega^2$, which leads to a correction of order $\omega^2/\mu^2$ compared to the zeroth-order term. Taken together, the maximal divergence arises from the fourth term, and its degree of divergence matches that of the zeroth-order contribution.

A similar analysis shows that higher–order terms exhibit the same maximal divergence, scaling as $\prod_{a=1}^{r}(\omega^2-\Omega_a^2)^{-1}$, with the even $k$-th order contribution formally ``suppressed" by a factor of order $(\omega^k/\mu^k)$.

We therefore conclude that, in the large angular momentum limit where $\Omega_a\to\omega$, the equilibrium partition function takes the form
\begin{align}\label{logz3}
\log Z \approx \frac{\mu^dF(\mu/T,\omega/T)}{\prod_{a=1}^{r}(\omega^2-\Omega_a^2)} .
\end{align}
This result closely parallels that obtained from fluid dynamics in the previous section. One important difference is that the undetermined function $F$ now depends on two dimensionless parameters, $\mu/T$ and $\omega/T$, in contrast to the fluid-dynamical regime where it depends on a single parameter. Because of this additional dependence, the predictive power is weaker than in fluid dynamics, and the behavior is therefore called semi-universal.

Although this expression was derived using a derivative expansion, we expect the functional form to be universal even beyond the regime in which the derivative expansion is strictly valid. Indeed, in the following Sections \ref{ex1} and \ref{ex2} we show that this structure persists in free theories and in the zero-temperature superfluid.

\subsubsection{Thermodynamic quantities}\label{thermo qu}

Starting from \eqref{logz3}, the thermodynamic quantities can be obtained by differentiating the free energy with respect to the corresponding chemical potentials. We employ the standard thermodynamic relations, as in \eqref{cons charge},
\begin{align}
J_a &= T\frac{\partial \log Z}{\partial \Omega_a},\qquad
Q = T\frac{\partial \log Z}{\partial\mu},\nonumber\\
S &= \log Z + T\frac{\partial \log Z}{\partial T},\qquad
E = T^2\frac{\partial \log Z}{\partial T}
+\mu Q
+\sum_{a=1}^{r}\Omega_aJ_a.
\end{align}

Now we express the microcanonical entropy in terms of charges using a procedure analogous to the one described \cite{Anand:2025mfh}. We can perform a saddle-point analysis in the limit where the angular velocities approach their critical value $\Omega_a \to \omega$.
Let's define the small parameters $\epsilon_a = \omega - \Omega_a$. In the limit $\Omega_a \to \omega$, we can approximate $\omega^2 - \Omega_a^2 \approx 2\omega\epsilon_a$. The grand canonical partition function then develops simple poles in $\epsilon_a$:
\begin{align}
\log Z \approx \frac{\mu^d F(\mu/T, \omega/T)}{(2\omega)^r \prod_{a=1}^r \epsilon_a} \equiv \frac{H(T, \mu)}{\prod_{a=1}^r \epsilon_a}
\end{align}
where $H(T,\mu)$ collects all regular, non-divergent contributions, and we set $\omega=1$ for convenience. The dependence on $\omega$ can be restored by dimensional analysis.

Using the thermodynamic relation for the angular momenta:
\begin{align}
J_a = T\frac{\partial \log Z}{\partial \Omega_a} = -T\frac{\partial \log Z}{\partial \epsilon_a} = T\frac{\log Z}{\epsilon_a} \implies \epsilon_a = \frac{T \log Z}{J_a}
\end{align}
Multiplying these parameters across all $r$ rotational planes gives:
\begin{align}
\prod_{a=1}^r \epsilon_a = \frac{T^r (\log Z)^r}{\prod_{a=1}^r J_a}
\end{align}
Equating this with $\prod_{a=1}^r \epsilon_a = \frac{H(T, \mu)}{\log Z}$ from our partition function assumption, we can solve for $\log Z$:
\begin{align}
\frac{T^r (\log Z)^r}{\prod_{a=1}^r J_a} = \frac{H(T, \mu)}{\log Z} \implies \log Z = \left( \frac{H(T, \mu)}{T^r} \right)^{\frac{1}{r+1}} \left( \prod_{a=1}^r J_a \right)^{\frac{1}{r+1}}
\end{align}

Let's evaluate the behavior of the remaining conserved charges. The electric/flavor charge $Q$ is given by:
\begin{align}
Q = T\frac{\partial \log Z}{\partial \mu} = \left(T \frac{\partial \log H}{\partial \mu}\right) \log Z
\end{align}
For the energy $E$, we introduce the ``twist" variable $\tau$ to capture the combination of energy and angular momenta that stays finite or scales slower than $J_a$, exactly as done in the paper:
\begin{align}
\tau \equiv E - \omega \sum_{a=1}^r J_a
\end{align}
Substituting the thermodynamic relation for $E$:
\begin{align}
\tau = T^2\frac{\partial \log Z}{\partial T} + \mu Q - \sum_{a=1}^r \epsilon_a J_a
\end{align}
We previously found that $\epsilon_a J_a = T \log Z$, meaning the last sum is simply $r T \log Z$. Plugging this in, along with our expression for $Q$, yields:
\begin{align}
\tau = \left( T^2 \frac{\partial \log H}{\partial T} + \mu T \frac{\partial \log H}{\partial \mu} - r T \right) \log Z
\end{align}

Finally, we compute the entropy using your provided relation:
\begin{align}
S = \log Z + T\frac{\partial \log Z}{\partial T} = \left( 1 + T \frac{\partial \log H}{\partial T} \right) \log Z
\end{align}
Notice a strict pattern here. If we define a collective scaling parameter $J_{\text{scale}} = \left( \prod_{a=1}^r J_a \right)^{\frac{1}{r+1}}$, then every thermodynamic quantity we derived—$\log Z$, $Q$, $\tau$, and $S$—is strictly proportional to $J_{\text{scale}}$, multiplied by some function of only $T$ and $\mu$:
\begin{align}
Q = \mathcal{Q}(T, \mu) J_{\text{scale}}, \quad \tau = \mathcal{T}(T, \mu) J_{\text{scale}}, \quad S = \mathcal{S}(T, \mu) J_{\text{scale}}
\end{align}
In the microcanonical ensemble, we invert the relations for $Q$ and $\tau$ to find $T$ and $\mu$ as functions of the intensive densities $\frac{Q}{J_{\text{scale}}}$ and $\frac{\tau}{J_{\text{scale}}}$. Substituting those back into the expression for $S$ produces an intensive entropy density function $s_{\text{int}}$.
This forces the microcanonical entropy to take exactly the semi-universal ``extensive" form:
\begin{align}
S(\tau, Q, J_a) \approx \left( \prod_{a=1}^r J_a \right)^{\frac{1}{r+1}} s_{\text{int}}\left( \frac{\tau}{\left( \prod_{a=1}^r J_a \right)^{\frac{1}{r+1}}}, \frac{Q}{\left( \prod_{a=1}^r J_a \right)^{\frac{1}{r+1}}} \right)
\end{align}
In this expression, the product $\left( \prod_{a=1}^r J_a \right)^{\frac{1}{r+1}}$ acts like a spatial volume, and $s_{\text{int}}$ acts as a purely dynamic, theory-dependent entropy density function.\footnote{One can also express the entropy as \begin{align}
S(\tau, Q, J_a) \approx Q \tilde{s}_{\text{int}}\left( \frac{\tau}{\left( \prod_{a=1}^r J_a \right)^{\frac{1}{r+1}}}, \frac{Q}{\left( \prod_{a=1}^r J_a \right)^{\frac{1}{r+1}}} \right)
\end{align}
since $Q$ also scales in the same way as $J$ and $\tau$.}
This formula characterizes the number of states—or, from the CFT perspective, the number of operators.

\subsubsection{Bound on twist}
In a relativistic CFT with radial quantization, the twist is defined as $\tau = E - J$, namely the difference between the scaling dimension (energy) and the angular momentum. This quantity plays an important role in the large-spin limit, as it measures the “distance” of a state from $E=J$. While unitarity of conformal representations alone does not strictly imply the non-negativity of the twist, additional physical constraints—most notably the Average Null Energy Condition (ANEC)—suggest that it is bounded from below \cite{Cordova:2017dhq}. This lower bound ensures, in particular, the convergence of the partition function as angular velocities approach their maximal values.

It is natural to ask what the appropriate analogue of the twist in a non-relativistic CFT is. For an NRCFT quantized in a harmonic trap, we define the corresponding quantity as a twist
\begin{align}
    \tau = E - \omega J =\omega(\Delta-J).
\end{align}
It turns out that the representation-theoretic unitarity bound by itself does not imply $\tau\ge0$. See Appendix~\ref{unitary} for details.

However, from \eqref{cons charge}, one can see that the twist for the non-relativistic fluid is positive:
\begin{align}
\tau=E-\omega J
&\approx
T^{2}\partial_T \log Z
+\mu T\partial_\mu \log Z
-2T \log Z\sum_{a=1}^{r}\frac{\Omega_a}{\omega+\Omega_a}\nonumber\\
&\approx_{\Omega_a\to\omega} (d-r)T\mu^{d}
\frac{F_0(\mu/T)}{\prod_{a=1}^r(\omega^2-\Omega_a^2)}>0
\end{align}
because $d>r$. 
Also, as we will see in Section \ref{ex1}, free theories satisfy this bound. This suggests that, for physical systems, it is natural for the twist $\tau$  to be at least bounded from below. However, we do not currently have a proof that an ANEC-like condition enforces $\tau\geq0$.
For a generic function $F$ in \eqref{logz3}, in \eqref{logz3}, the positivity of the twist is not guaranteed. Therefore, if a twist-positivity condition holds in a unitary NRCFT, it would impose constraints on the function $F$.

\section{Example: Free partition function}\label{ex1}

In this section and the next, we consider examples that do not (necessarily) fit within a fluid-dynamical description, but nevertheless exhibit semi-universal behavior at large angular momentum.

First, we consider a free nonrelativistic system in a harmonic trap, working in the grand canonical ensemble
\begin{align}
Z = \mathrm{Tr}\,\exp\!\left[-\beta\left(H-\mu N-\sum_a\Omega_a J_a\right)\right],
\end{align}
where $\mu$ is the chemical potential and $\Omega$ is the angular velocity conjugate to the angular momentum $J$.
In the isotropic harmonic trap in $d$ spatial dimensions, the single-particle spectrum is given by
\begin{align}
\varepsilon_{\vec n}
=
\omega\left(\sum_{i=1}^d n_i + \frac d2\right),
\qquad
n_i=0,1,2,\dots
\end{align}

\subsection{Bosons}

Defining
\begin{align}
N=\sum_{i=1}^d n_i ,
\end{align}
the energy levels and degeneracies are
\begin{align}
\varepsilon_N=\omega\left(N+\frac d2\right),
\qquad
g_d(N)=\binom{N+d-1}{d-1}.
\end{align}

In the following, we subtract the zero-point energy by a shift of $\mu$, so that the lowest single-particle energy is set to zero.

The exact grand canonical partition function for free bosons is
\begin{align}
\log Z_{\rm bos}
=
-\sum_{N=0}^\infty
\binom{N+d-1}{d-1}
\log\!\left(1-e^{-\beta(\omega N-\mu)}\right).
\end{align}

Using the expansion
\begin{align}
-\log(1-q)=\sum_{k=1}^\infty \frac{q^k}{k},
\end{align}
and performing the sum over $N$, one obtains the exact closed-form expression
\begin{align}
\log Z_{\rm bos}
=
\sum_{k=1}^\infty
\frac{e^{k\beta\mu}}{k}
\frac{1}{\left(1-e^{-k\beta\omega}\right)^d}
\end{align}
which is valid for all integer $d$ and $\mu\le 0$.

We now include rotation along the maximal Cartan subalgebra of $SO(d)$.
Let
\begin{align}
r=\left\lfloor \frac d2 \right\rfloor
\end{align}
be the rank of $SO(d)$, and introduce independent angular velocities
$\Omega_a$ ($a=1,\dots,r$), each conjugate to a Cartan generator $J_a$.
The grand canonical ensemble is defined by
\begin{align}
Z=\mathrm{Tr}\,
\exp\!\left[-\beta\left(H-\mu N-\sum_{a=1}^r \Omega_a J_a\right)\right].
\end{align}

For an isotropic harmonic trap, the single-particle Hamiltonian decomposes
into $r$ orthogonal rotation planes with frequencies
\begin{align}
\omega_{a,\pm}=\omega\pm\Omega_a ,
\qquad a=1,\dots,r,
\end{align}
while, for odd $d$, the remaining one direction is non-rotating with
frequency $\omega$.

Repeating the steps of the non-rotating computation, one finds the exact
bosonic grand partition function
\begin{align}
\log Z_{\rm bos}
=
\sum_{k=1}^\infty
\frac{e^{k\beta\mu}}{k}
\prod_{a=1}^r
\frac{1}{\left(1-e^{-k\beta(\omega-\Omega_a)}\right)
\left(1-e^{-k\beta(\omega+\Omega_a)}\right)}
\times
\begin{cases}
1, & d=2r,\\[6pt]
\displaystyle \frac{1}{\left(1-e^{-k\beta\omega}\right)}, & d=2r+1.
\end{cases}
\end{align}
The stability condition requires
\begin{align}
|\Omega_a|<\omega,
\qquad a=1,\dots,r.
\end{align}
Expanding at large spin limit $\beta(\omega-\Omega) \ll 1$ reproduces 
\begin{align}
    \log Z\approx \sum_{n=1}^{\infty}\frac{(2\beta\omega)^r}{\omega^2-\Omega_a^2}\tilde{h}_{\rm bos}(\beta\mu,\beta\omega)+\cdots
\end{align}
where
\begin{align}
    \tilde{h}_{\rm bos}(\beta\mu,\beta\omega)=\sum_{k=1}^{\infty}\frac{e^{k\beta\mu}}{k}\prod_{a=1}^r
\frac{1}{
\left(1-e^{-2k\beta\omega}\right)}
\times
\begin{cases}
1, & d=2r,\\[6pt]
\displaystyle \frac{1}{\left(1-e^{-k\beta\omega}\right)}, & d=2r+1.
\end{cases}
\end{align}

\subsection{Fermions}

For free fermions, the exact grand partition function differs only by the statistics. Starting from
\begin{align}
\log Z_{\rm ferm}
=
\sum_{\vec n}
\log\!\left(1+e^{-\beta(\varepsilon_{\vec n}-\mu)}\right),
\end{align}
one finds, after the same steps,
\begin{align}
\log Z_{\rm ferm}
=
\sum_{k=1}^\infty
\frac{(-1)^{k+1}e^{k\beta\mu}}{k}
\frac{1}{\left(1-e^{-k\beta\omega}\right)^d}
\end{align}
in the absence of rotation.

Including rotation, the exact fermionic partition function is
\begin{align}
\log Z_{\rm ferm}
=
\sum_{k=1}^\infty
\frac{(-1)^{k+1}e^{k\beta\mu}}{k}
\prod_{a=1}^r
\frac{1}{\left(1-e^{-k\beta(\omega-\Omega_a)}\right)
\left(1-e^{-k\beta(\omega+\Omega_a)}\right)}
\times
\begin{cases}
1, & d=2r,\\[6pt]
\displaystyle \frac{1}{\left(1-e^{-k\beta\omega}\right)}, & d=2r+1.
\end{cases}
\end{align}
which is well defined for $|\Omega|<\omega$.
Expanding at large spin limit $\beta(\omega-\Omega) \ll 1$ reproduces 
\begin{align}
    \log Z\approx \sum_{n=1}^{\infty}\frac{(2\beta\omega)^r}{\omega^2-\Omega_a^2}\tilde{h}_{\rm ferm}(\beta\mu,\beta\omega)+\cdots
\end{align}
where
\begin{align}
    \tilde{h}_{\rm ferm}(\beta\mu,\beta\omega)=\sum_{k=1}^{\infty}\frac{(-1)^{k+1}e^{k\beta\mu}}{k}\prod_{a=1}^r
\frac{1}{
\left(1-e^{-2k\beta\omega}\right)}
\times
\begin{cases}
1, & d=2r,\\[6pt]
\displaystyle \frac{1}{\left(1-e^{-k\beta\omega}\right)}, & d=2r+1.
\end{cases}
\end{align}

\section{Example: Cold Atom}\label{ex2}

In this section, we focus on superfluid states in conformal field theories. At sufficiently low temperature a many-body system can undergo Bose condensation and enter a superfluid phase. A representative physical example is a gas of fermions at unitarity \cite{Son:2008ye}, which realizes a strongly interacting non-relativistic conformal system but is notoriously difficult to analyze directly using microscopic methods.

A useful simplification arises at large charge (equivalently, large particle number). In this regime the ground state admits a controlled effective field theory description, allowing analytic access to equilibrium properties. Large-charge effective field theories have been extensively developed in relativistic CFTs \cite{Hellerman:2015nra,Alvarez-Gaume:2016vff,Cuomo:2019ejv,Cuomo:2022kio,Cuomo:2023mxg,Badel:2022fya,Dondi:2024vua,Choi:2025tql,Lee:2025qim}, and the same philosophy applies to non-relativistic conformal systems.

In this section we review large-charge superfluids in NRCFTs. Most of the material is a summary of known results, but the analysis of superfluids at large angular momentum—specifically in the regime \(\mu(\omega-\Omega_a)\to 0\)—contains new results.\footnote{Because the superfluid EFT is formulated in real time rather than in thermal time, it describes the zero-temperature ground state and its excitations. Since $T=0$, the relevant dimensionful parameters are the chemical potential \(\mu\) and the trapping frequency \(\omega\), and the large--angular-momentum limit is characterized by \(\mu(\omega-\Omega_a)\to 0\), as we will see later in this section.}
For detailed background explanataions, we refer \cite{Cooper_2008,Fetter:2009zz,Geracie:2014nka,Kravec:2018qnu,Kravec:2019djc,Pellizzani:2024gfa} and references therein.

\subsection{Review on large charge superfluid in NRCFT}
The leading-order effective Lagrangian describing a nonrelativistic superfluid coupled to an external potential $A_0$ is given by \cite{Son:2005rv}
\begin{align}\label{lag}
    \mathcal{L} = P(X),
    \quad
    X \equiv \partial_0 \chi - A_0 - \frac{1}{2} (\partial_i \chi)^2.
\end{align}
where $\chi$ is the phase of the condensate (or the Goldstone boson associated with the spontaneously broken U(1) symmetry).
In the special case of an NRCFT one has 
\begin{align}
    \mathcal{L} = c_0\,X^{\frac{d+2}{2}},
    \quad
    A_0 = \tfrac12\,\omega^2 r^2
    \quad(\text{e.g.\ fermions at unitarity}).
\end{align}

The charge density and superfluid velocity follow directly from \eqref{lag}:
\begin{align}
    n = \frac{\partial \mathcal{L}}{\partial \dot\chi}
      = c_0\Bigl(\tfrac{d}{2}+1\Bigr) X^{\frac{d}{2}},
    \qquad
    v_i = -\,\partial_i \chi.
\end{align}
Accordingly, the charge current is
\begin{align}
    j^{\mu} = \bigl(n,\,n\,v^i\bigr),
\end{align}
and its conservation, $\partial_\mu j^\mu = 0,$ is equivalent to the equation of motion.
The Hamiltonian density then reads
\begin{align}
    \mathcal{H}
    &= n\,\dot\chi - \mathcal{L}
      = n\Bigl(X + A_0 + \tfrac12\,v^2\Bigr) - P(X)
    \nonumber\\
    &= \tfrac12\,n\,v^2 + \epsilon(n) + n\,A_0,
    \quad
    \epsilon(n) \equiv n\,X - P(X) = \tfrac{d}{2}\,\mathcal{L}.
\end{align}

To find the lowest‐energy configuration at fixed charge, one sets
\begin{align}
    \chi = \mu\,t,
\end{align}
which solves the equation of motion, and gives
\[
    n = c_0\Bigl(\tfrac{d}{2}+1\Bigr)\bigl(\mu - A_0\bigr)^{\tfrac{d}{2}}.
\]
For \(A_0 = \tfrac12\,\omega^2 r^2\), the density decreases radially and classically vanishes at the Thomas–Fermi radius $R_{\rm TF}$ 
\[
    R_{\rm TF} = \sqrt{\frac{2\mu}{\omega}}.
\]
Therefore, the total charge scales as
\(\,Q\propto (\mu/\omega)^{d}\) and the energy as
\(\,E_Q\propto Q^{\frac{d+1}{d}}\).

One can study fluctuations around the homogeneous superfluid background. To this end, write $\chi_{cl}=\mu t + \pi$; solving the equations of motion gives the quadratic phonon mode
\begin{align}
    \mathcal{H}_{ph}=\sum_{n,l}\omega(n,l)\pi_{n,l}^{\dagger}\pi_{n,l}
\end{align}
where
\begin{align}
    \omega(n,l)=\omega\left(\frac{4}{d}n^2+(4-\frac{4}{d})n+\frac{4}{d}nl+l\right)^{1/2}.
\end{align}
This determines the excitation spectrum as a function of the charge and the angular momentum.

As the angular momentum increases, the wave function of the surface modes becomes increasingly localized near the edge of the trap, and the phonon effective field theory ceases to be valid. For example, in $d=3$, the phonon description breaks down when the angular momentum reaches $J\sim Q^{1/3}$ (see \cite{Kravec:2019djc} for details).

\subsection{Few vortices}
As the angular momentum is increased beyond the regime of validity of the phonon effective field theory, the low-energy description in terms of smooth Goldstone fluctuations breaks down. Physically, this occurs because the angular momentum is no longer efficiently carried by delocalized phonon modes: instead, the system lowers its energy by forming localized topological excitations. These new degrees of freedom are vortices, which become the dominant carriers of angular momentum at sufficiently large $J$.

A vortex is defined as a configuration of the superfluid phase field $\chi$ for which the superfluid velocity
\[
v_i \equiv \partial_i \chi
\]
is irrotational everywhere except at codimension-2 defects, known as vortex cores. At the location of a vortex core, the phase field becomes singular and the circulation is quantized. For a vortex with integer vorticity $q$, the phase winds according to
\[
\oint d\chi = 2\pi q ,
\]
where the integral is taken along a closed contour encircling the vortex core. The quantization of circulation follows from the single-valuedness of the superfluid order parameter and is a robust, topological property of the superfluid state.

From now on, let us consider $d=2$ for simplicity such that a vortex is a point-like object. (For higher dimensions vortices are higher dimensional objects. Generalization is straightforward and we present results at the end of the section.)

In the presence of a single vortex centered at $\mathbf{R}$, the leading‐order velocity profile is
\begin{align}
    v^i = -\,\partial^i \chi
         = \frac{\epsilon^{ij}\,(r_j - R_j)}{|\mathbf{r} - \mathbf{R}|^2}.
\end{align}

One can set the vortex core radius $a$ by requiring that the local effective chemical potential is balanced by the interaction potential between vortices:
\begin{align}
    \mu_{\rm eff}(r) \equiv \mu - \frac{\omega^2 r^2}{2}
      \sim \frac{1}{2|\mathbf{r}-\mathbf{R}|^2}.
\end{align}
In the bulk one finds $a\sim 1/\sqrt{\mu_{\rm eff}}\sim 1/\sqrt{\mu}$. Near the edge of the superfluid the vortex core size increases, but its contribution is negligible from the viewpoint of the overall extensive properties. The self‐energy of a single vortex scales as
\[
    E_{\rm self} \approx \int \frac{nv^2}{2} \sim \mu\,\log\mu.
\]
where the integration domain is given by $|r-R|>a, ~r<R_{TF}$. (UV and IR cutoff)

Suppose that there is one vortex at the center.
The angular momentum contribution from the vortex is given by $J\approx 2\pi c_0\frac{\mu^2}{\omega^2}$ \cite{Kravec:2019djc}.
Therefore, angular momentum and the excitation energy relation is given as \cite{Kravec:2019djc}
\begin{align}
    \delta E\approx\sqrt{\frac{c_0\pi}{2}}\sqrt{J}\log J
\end{align}

One can determine the vortex core radius \(a\) by requiring that the local effective chemical potential is balanced by the interaction potential between vortices:
\begin{align}
    \mu_{\rm eff}(r)\equiv \mu-\frac{\omega^2 r^2}{2}
    \sim \frac{1}{2|\mathbf r-\mathbf R|^2}.
\end{align}
In the bulk this gives \(a\sim 1/\sqrt{\mu_{\rm eff}}\sim 1/\sqrt{\mu}\). 
The self-energy of a single vortex scales as
\begin{align}
    E_{\rm self}\approx \int \frac{n v^2}{2}\sim \mu\,\log\mu,
\end{align}
where the integration domain is defined by the ultraviolet and infrared cutoffs,
\(|\mathbf r-\mathbf R|>a\) and \(r<R_{\rm TF}\), respectively.

\subsection{Rotating vortices and large spin limit}

When the total number of vortices in the system becomes much larger than one, the vortex density \(n\) may be treated as a continuous function on the scale of the system size. To determine the ground state, we minimize the energy at fixed angular momentum \(J\) and charge \(Q\) using Lagrange multipliers:
\begin{align}\label{multiplier}
\int d^dx \left(\frac{1}{2}n v^2+\epsilon(n)+nA_0\right)
-\Omega\left(\int d^dx\, n(r\times v)\cdot \hat z -J\right)
-\mu\left(\int d^dx\, n-Q\right).
\end{align}
The first term is the energy functional, while the second and third terms impose the constraints on \(J\) and \(Q\), respectively. Varying with respect to \(v\) (keeping \(n\) fixed) yields
\begin{align}
v=\Omega\,\hat{z}\times r,
\end{align}
which shows that, in a coarse-grained sense, the superfluid undergoes rigid rotation with angular velocity \(\Omega\). Varying with respect to \(n\) (keeping \(v\) fixed) gives
\begin{align}\label{cdensity}
    n\propto \left(\mu-\frac{1}{2}(\omega^2-\Omega^2)r^2\right)^{d/2}.
\end{align}
Note that \(\Omega<\omega\) is required to confine the superfluid, i.e. there is an upper bound on the allowed angular velocity.

The total charge is
\begin{align}
    Q=\int d^dx n \propto \left(\frac{\mu}{\sqrt{\omega^2-\Omega^2}}\right)^{d}
\end{align}
where the cloud has a finite radius set by the point where the density vanishes $R_{\rm TF}=\sqrt\frac{2\mu}{\omega^2-\Omega^2}$
The angular momentum is
\begin{align}
    J=\int d^dx nr^2\Omega\propto \Omega\int dr r^{d+1}(\mu-\frac{1}{2}(\omega^2-\Omega^2)r^2)^{d/2}
\end{align}

Let us specialize to spatial dimension \(d=2\), for which
\begin{align}
    n=2c_0X,\quad~X=\mu-\frac{1}{2}(\omega^2-\Omega^2)r^2,
\end{align}
This is equivalently expressed as \(\partial_0\chi=\mu+\Omega^2 r^2\) and \(|\partial_i\chi|=\Omega r\).
The energy follows from the Hamiltonian density,
\begin{align}
    E=\int 2\pi r dr \mathcal{H}= \frac{4 \pi c_0  \mu ^3 \omega ^2}{3 \left(\omega ^2-\Omega ^2\right)^2},
\end{align}
since $\mathcal{H}=\frac{1}{2}n\left(r^2(\frac{1}{2}\omega^2+\frac{3}{2}\Omega^2)+\mu\right)$. Similarly, the charge and angular momentum are
\begin{align}\label{chargeQJ}
    Q= \frac{2\pi c_0\mu^2}{\omega^2-\Omega^2},\qquad
    J= \frac{4\pi c_0\mu^3\Omega}{3(\omega^2-\Omega^2)^2}.
\end{align}
The quantities \(E\), \(J\), and \(Q\) satisfy the dispersion relation
\begin{align}\label{relations}
    \left(\frac{E}{\omega}\right)^2-J^2=\frac{2}{9\pi c_0}Q^3.
\end{align}
From these expressions one obtains the leading-order partition function,
\begin{align}
    \frac{1}{\beta}\log Z
    =\frac{2\pi c_0}{3}\frac{\mu^3}{(\omega^2-\Omega^2)}.
\end{align}
which suits well with the expectation given in \eqref{logz3}. Note that since $T=0$, the only possible scaling is $\frac{1}{\beta}\log Z \sim\mu^{d-1}$.

\paragraph{Regime of Validity}
The EFT analysis in this subsection is valid when the characteristic size of the system is much larger than the typical vortex separation, so that the vortex density may be approximated as a continuum. In addition, the vortex separation must be much larger than the EFT cutoff length, ensuring that the vortex self-energy gives a negligible contribution to the total energy.

The vortex density is given as
\begin{align}
    n_{v}= \frac{1}{2\pi}(\partial_1v_2-\partial_1v_2),
\end{align}
so that $\int_{A}d^2x n_{v}=\frac{1}{2\pi} \oint_C v\cdot dl=\frac{1}{2\pi}\oint d\chi=N_v$.
Since $(v_1,v_2)=\Omega r(-\sin\theta,\cos\theta)=\Omega(-y,x)$, 
\begin{align}\label{vdensity}
    n_v = \frac{\Omega}{\pi}.
\end{align}
Thus, the vortex density is constant, unlike the charge density.

The EFT cutoff length scales as \(\sim \mu^{-1/2}\), the vortex–vortex separation as \(\sim n_v^{-1/2}\sim \Omega^{-1/2}\), and the Thomas–Fermi radius of the cloud as \(R_{\rm TF}\sim \sqrt{\mu}/\omega\). Therefore, the regime of validity of the EFT requires
\begin{align}\label{criteria}
\frac{\sqrt{\mu}}{\omega}\gg \frac{1}{\sqrt{\Omega}}\gg\frac{1}{\sqrt{\mu}},
\end{align}
with \(\omega>\Omega\).
Remarkably, this regime also coincides with the regime in which the derivative expansion is parametrically controlled for the holographic theories discussed in Section \ref{fluid} and in footnote $4$ in Subsection \ref{3}, in the zero-temperature limit.

\subsection{Vortex lattice and beyond}

Starting from the rotating vortex solution described in the previous subsection, we now consider increasing the angular momentum while keeping the total charge fixed. This requires increasing \(\Omega\) while decreasing \(\mu\).

In the limit $\Omega \to \omega$ (while maintaining $\mu \gg \omega$), the system enters a vortex-lattice phase. Strictly speaking, the lattice structure is only approximate, since the rotation frequency must remain slightly below the trapping frequency ($\Omega = \omega - \epsilon$); otherwise, the centrifugal force overcomes the harmonic confinement, making the system unbound. In this near-critical regime, the system's characteristic size is defined by the Thomas-Fermi radius, $R_{\text{TF}} \propto \sqrt{\mu/(\omega^2-\Omega^2)}$. Consequently, thermodynamic quantities scale with an effective volume factor $V \sim \left[\mu/(\omega^2-\Omega^2)\right]^{d/2}$. The divergence of this volume factor as $\Omega \to \omega$ is the essential source of the semi-universal thermodynamic scaling.

Within the bulk of this expanded cloud, both the vortex density \eqref{vdensity} and the charge density \eqref{cdensity} become locally uniform, approximating a translationally invariant lattice system. Let us emphasize again that the large-charge EFT is valid as long as $\mu\gg\omega$.

Vortex lattices in NRCFTs were originally discussed in \cite{Son:2005ak}, and further developed in subsequent studies \cite{Watanabe:2013iia, Moroz:2018noc} using the Gross-Pitaevskii (GP) Lagrangian. The GP framework is essentially equivalent to the large-charge EFT in the regime where the latter is valid. For $d=2$, the GP Lagrangian is given by:
\begin{align}
    \mathcal{L}=\frac{i}{2}(\psi^{\dagger}\dot{\psi}-\dot{\psi}^{\dagger}\psi)-\frac{1}{2m}|\nabla\psi|^2-V_{\text{trap}}(\vec{x})\psi^{\dagger}\psi-\frac{1}{2}g(\psi^{\dagger}\psi)^2.
\end{align}

Parameterizing the field in terms of density and phase as $\psi=\sqrt{n}e^{-i\theta}$, we can isolate the Thomas-Fermi regime where the quantum pressure term is negligible compared to the interaction energy ($\frac{1}{m}|\nabla \sqrt{n}|^2 \ll g n^2$). Under this condition, the equation of motion for the density field $n$ simplifies, and the action maps directly onto that of the large-charge EFT \eqref{lag} \cite{Cuomo:2023vvd}. To minimize its energy, the system crystallizes into an Abrikosov-like configuration; as demonstrated by Tkachenko, the stability of the rotating superfluid is maximized in a triangular lattice \cite{Tkachenko1965}.

As explained, this vortex lattice description remains robust as long as $\mu \gg \omega$. In this regime, the angular momentum and the particle number satisfy the scaling relation $Q^{3/2} \ll J \ll Q^2$, consistent with equation \eqref{chargeQJ}.

As we continue to push the system toward the limit $J \sim Q^2$, the chemical potential drops to $\mu \sim \mathcal{O}(\omega)$, and the large-charge EFT description begins to falter. The vortex cores expand and begin to overlap, driving a phase transition into a Fractional Quantum Hall (FQH) phase.

The mechanics of this transition are best understood from the GP perspective. In the rotating frame, the Coriolis force creates a synthetic magnetic field with an effective cyclotron frequency $2\Omega$. As $\mu$ decreases, the overall density $n$ drops, which effectively suppresses the relative strength of the nonlinear interaction term $gn^2$. As the filling factor $\nu \equiv n/n_{v}$ (the ratio of particle density to vortex density) approaches order unity, the system is essentially restricted to the Lowest Landau Level (LLL).

While it is superficially counterintuitive that a weakly interacting system leads to a strongly correlated FQH state, this is a direct consequence of the massive degeneracy of the LLL. Because kinetic energy is quenched in the LLL, even a perturbatively small interaction $g$ becomes the dominant energy scale. The classical GP mean-field approximation—which assumes particles can be described by a single macroscopic coherent wavefunction $\psi$—breaks down entirely. Instead, the system enters a Bosonic FQH state (such as the $\nu=1/2$ Laughlin state), where particles avoid each other through complex many-body correlations. As noted in the literature, for filling factors smaller than a critical value $\nu_c$, the vortex lattice phase is unstable to quantum fluctuations and is replaced by a strongly correlated phase. (See \cite{Cooper_2008} and references therein.)

Finally, if the angular momentum is increased even further such that $\nu \ll 1$ (corresponding to $\mu \ll \omega$), the system enters a limit that serves as the non-relativistic analogue of the Regge limit \cite{Alday:2007mf,Komargodski:2012ek,Fitzpatrick:2012yx}. In this extreme high-spin regime, the particles become so dilute that the interaction energy vanishes ($V_{\text{int}} \to 0$). The physics is completely dominated by the single-particle harmonic oscillator modes of the LLL. The anomalous dimensions of the operators vanish, and the system behaves as a collection of non-interacting particles in high-angular-momentum states, reflecting the linearity of the Regge trajectory. As described in Section \ref{ex1}, the equilibrium partition function in this free-theory regime still exhibits semi-universality, provided the effective temperature remains high compared to the gap, $\beta(\omega-\Omega)\ll 1$.
Since $\mu\gg \omega$ and $\mu\ll \omega$ exhibit semi-universality, it is natural to expect that the FQH case with 
$\mu\sim\omega$ also follows the same behavior.

\section{Discussion}\label{dis}

In this paper, we argued that the thermodynamic equilibrium partition function of an NRCFT at large angular momentum takes the universal form given in \eqref{logz3}.
Morally, the divergent factor arises because the effective volume of the system increases as $\Omega$ approaches $\omega$: the centrifugal force and the trapping force balance each other, allowing the matter to spread over a larger region.

Throughout the paper, we assumed parity invariance. It would be interesting to extend the analysis to parity-violating theories (for example those coupled to Chern--Simons terms). We expect that the semi-universal pole structure will persist, although the detailed form of the residue may be modified.

\paragraph{Implications for holographic NRCFTs.}
In a holographic realization of an NRCFT, it would be valuable to identify a rotating black hole solution in global Schrödinger geometry and test whether the predicted semi-universal behavior is reproduced at large angular momentum. Although no explicit solution is currently known, one may speculate about the qualitative thermodynamic structure. 

Suppose there exists a branch of rotating black holes that becomes thermodynamically unstable (i.e.\ has negative specific heat) below a certain critical size, terminating at a minimal temperature \(\beta=\beta_*\). By analogy with the relativistic analysis of \cite{Anand:2025mfh}, one may then examine the resulting phase diagram and in particular the possible existence of an intermediate ``grey galaxy'' phase interpolating between a large black hole and a thermal gas.

In this regime, as the temperature approaches the critical value $\beta \to \beta_*$, the angular velocity $\Omega$ must simultaneously approach its critical value $\omega$. We define the small deviation parameters:
\begin{align}
    \delta \beta = \beta - \beta_*, \quad \nu = \omega^2 - \Omega^2.
\end{align}
We assume the regime where the temperature deviation dominates the chemical potential scale, $\delta\beta \gg \nu$. Based on the scaling arguments for critical phenomena in this geometry, we propose that the on-shell action (logarithm of the partition function) for the black hole takes the form:
\begin{align}
    \log Z_{\rm BH} \approx \frac{A}{G_N} \frac{(\delta\beta)^{r+1}}{\nu^r},
    \label{eq:logZ_BH_ansatz}
\end{align}
where $A$ is a numerical constant, $G_N$ is the Newton constant (suppressed in the ``large $N$" limit), and we consider the spatial dimension to be even $d=2r$ with equal angular velocities, $\Omega_a=\Omega$, for simplicity. 

The thermodynamic charges of the black hole—angular momentum $J$ and twist $\tau = E - \omega J$—are derived as conjugate variables to $\nu$ and $\delta\beta$:
\begin{align}
    J_{\rm BH} &= -\frac{\partial \log Z_{\rm BH}}{\partial \nu} \propto \frac{1}{G_N} \frac{(\delta\beta)^{r+1}}{\nu^{r+1}}, \\
    \tau_{\rm BH} &\approx -\frac{\partial \log Z_{\rm BH}}{\partial \beta} \propto \frac{1}{G_N} \frac{(\delta\beta)^r}{\nu^r}.
\end{align}
Eliminating the chemical potentials $\nu$ and $\delta\beta$, we find a state equation relating the conserved charges, independent of the temperature:
\begin{align}
    G_N J_{\rm BH} \propto (G_N \tau_{\rm BH})^{\frac{r+1}{r}}.
    \label{eq:BH_charge_relation}
\end{align}
This relation defines the ``Black Hole Phase" in the microcanonical ensemble.

However, the full system also includes the contribution from the thermal gas in the bulk. Unlike the black hole, the gas partition function is a one-loop quantum effect (order $G_N^0$) and is expected to have a simple pole structure as $\beta \to \beta_*$:
\begin{align}
    \log Z_{\rm gas} \approx \frac{B - C\delta\beta}{\nu^r},
    \label{eq:logZ_gas_ansatz}
\end{align}
where $B$ and $C$ are constants. Crucially, note the absence of the $1/G_N$ factor, indicating that the gas is thermodynamically subleading to the black hole when $\delta\beta$ is finite.

The ``Grey Galaxy" phase emerges in the regime where the black hole becomes small enough that its entropy is comparable to or smaller than the gas entropy, yet it is still in equilibrium with the gas. The total partition function is:
\begin{align}
    \log Z_{\rm tot} = \log Z_{\rm BH} + \log Z_{\rm gas}.
\end{align}
A phase transition occurs due to the different scaling of the two contributions with respect to $\delta\beta$. 
Comparing \eqref{eq:logZ_BH_ansatz} and \eqref{eq:logZ_gas_ansatz}, we see that as $\delta\beta \to 0$:
\begin{itemize}
    \item The Gas free energy scales as $(\delta\beta)^0 \sim \mathcal{O}(1)$.
    \item The Black Hole free energy scales as $(\delta\beta)^{r+1} \to 0$.
\end{itemize}
Since $r > 0$, the black hole contribution to the free energy vanishes faster than the gas contribution as we approach the minimal temperature. This suggests a continuous phase transition where the black hole evaporates into the gas phase. 

Specifically, for a fixed large angular momentum $J$, as we lower the twist $\tau$, the system moves from the pure black hole phase obeying \eqref{eq:BH_charge_relation} into the grey galaxy phase. In the grey galaxy phase, the chemical potentials are effectively fixed by the dominant gas saddle ($\nu \approx \text{const}$), while the black hole acts as a small, high-entropy defect. The critical exponent $r$ thus determines the universality class of this transition in the non-relativistic setting.


\section*{Acknowledgments}
We would like to thank Shiraz Minwalla for valuable discussions. We also thank Seok Kim and Jaehyeok Choi for discussions related to Section 6. The work of E.L. was supported by the Infosys Endowment for the study of the Quantum Structure of Spacetime.

\appendix

\section{Unitarity Bound}\label{unitary}

In this Appendix we clarify the relation between the Schrödinger unitarity bound and the positivity of the non-relativistic twist
\begin{align}
\tau \equiv E-\omega J .
\end{align}
In particular, we show that the representation-theoretic unitarity bound by itself does not imply $\tau\ge0$.  
The reason is that the trapped Hamiltonian naturally separates into a universal center-of-mass contribution, which is constrained by the symmetry algebra, and an internal contribution, which is not.

This separation follows directly from the central extension of the Schrödinger algebra,
\begin{align}
[K_i,P_j]=i\delta_{ij}N ,
\end{align}
where $N$ is the particle number.  
In any fixed particle-number sector $N>0$, 
one constructs the operators
\begin{align}
H_{\rm cm}=\frac{P^2}{2N}, \qquad C_{\rm cm}=\frac{K^2}{2N},
\end{align}
which obey the same commutation relations with $P_i$ and $K_i$ as the full generators $H$ and $C$:
\begin{align}
[H_{\rm cm},K_i]=-iP_i, \qquad [C_{\rm cm},P_i]=iK_i .
\end{align}
Consequently the differences
\begin{align}
H_{\rm int}\equiv H-H_{\rm cm}, \qquad
C_{\rm int}\equiv C-C_{\rm cm}
\end{align}
commute with the Heisenberg generators,
\begin{align}
[H_{\rm int},P_i]=[H_{\rm int},K_i]=0, \qquad
[C_{\rm int},P_i]=[C_{\rm int},K_i]=0 .
\end{align}
They therefore act only on $\mathcal H_{\rm int}$ and represent genuine internal degrees of freedom.

The trapped Hamiltonian splits as
\begin{align}
H_{\rm trap}=H+\omega^2 C
=\underbrace{\frac{P^2}{2N}+\omega^2\frac{K^2}{2N}}_{H_{\rm cm}^{\rm osc}}
+\;H_{\rm int}^{\rm trap},
\end{align}
where $H_{\rm int}^{\rm trap}=H_{\rm int}+\omega^2 C_{\rm int}$ commutes with $P_i$ and $K_i$.  
Similarly the angular momentum decomposes as
\begin{align}
J=J_{\rm cm}+J_{\rm int},
\end{align}
with $J_{\rm int}$ commuting with $P_i$ and $K_i$.

The center-of-mass part
\begin{align}
H_{\rm cm}^{\rm osc}=\frac{P^2}{2N}+\omega^2\frac{K^2}{2N}
\end{align}
is fixed entirely by the Heisenberg algebra.  
Introducing
\begin{align}
a_i=\frac{1}{\sqrt{2N\omega}}\left(P_i-i\omega K_i\right),
\end{align}
one finds $[a_i,a_j^\dagger]=\delta_{ij}$ and therefore
\begin{align}
H_{\rm cm}^{\rm osc}
=
\omega\left(\sum_i a_i^\dagger a_i+\frac{d}{2}\right),\qquad
J_{\rm cm}=\sum_{i<j} a_i^\dagger a_j - a_j^\dagger a_i .
\end{align}
Hence
\begin{align}
H_{\rm cm}^{\rm osc}-\omega J_{\rm cm}\ge0 .
\end{align}

The internal operators $H_{\rm int}^{\rm trap}$ and $J_{\rm int}$, however, commute with $P_i$ and $K_i$ and are not constrained by the oscillator algebra.  
There is therefore no general operator inequality
\begin{align}
H_{\rm int}^{\rm trap}-\omega J_{\rm int}\ge0
\end{align}
for arbitrary states.  
Consequently the full twist
\begin{align}
\tau=H_{\rm trap}-\omega J
\end{align}
is not positive on the entire Hilbert space, even though the center-of-mass sector always satisfies the bound.

\bibliography{References}

@article{Bhattacharyya:2007vs,
    author = "Bhattacharyya, Sayantani and Lahiri, Subhaneil and Loganayagam, R. and Minwalla, Shiraz",
    title = "{Large rotating AdS black holes from fluid mechanics}",
    eprint = "0708.1770",
    archivePrefix = "arXiv",
    primaryClass = "hep-th",
    doi = "10.1088/1126-6708/2008/09/054",
    journal = "JHEP",
    volume = "09",
    pages = "054",
    year = "2008"
}

@article{Benjamin:2023qsc,
    author = "Benjamin, Nathan and Lee, Jaeha and Ooguri, Hirosi and Simmons-Duffin, David",
    title = "{Universal asymptotics for high energy CFT data}",
    eprint = "2306.08031",
    archivePrefix = "arXiv",
    primaryClass = "hep-th",
    reportNumber = "CALT-TH 2023-014, IPMU 23-0020",
    doi = "10.1007/JHEP03(2024)115",
    journal = "JHEP",
    volume = "03",
    pages = "115",
    year = "2024"
}

@article{Kim:2023sig,
    author = "Kim, Seok and Kundu, Suman and Lee, Eunwoo and Lee, Jaeha and Minwalla, Shiraz and Patel, Chintan",
    title = "{Grey Galaxies\textquoteright{} as an endpoint of the Kerr-AdS superradiant instability}",
    eprint = "2305.08922",
    archivePrefix = "arXiv",
    primaryClass = "hep-th",
    doi = "10.1007/JHEP11(2023)024",
    journal = "JHEP",
    volume = "11",
    pages = "024",
    year = "2023"
}

@article{Cuomo:2022kio,
    author = "Cuomo, Gabriel and Komargodski, Zohar",
    title = "{Giant Vortices and the Regge Limit}",
    eprint = "2210.15694",
    archivePrefix = "arXiv",
    primaryClass = "hep-th",
    doi = "10.1007/JHEP01(2023)006",
    journal = "JHEP",
    volume = "01",
    pages = "006",
    year = "2023"
}

@article{Hellerman:2015nra,
    author = "Hellerman, Simeon and Orlando, Domenico and Reffert, Susanne and Watanabe, Masataka",
    title = "{On the CFT Operator Spectrum at Large Global Charge}",
    eprint = "1505.01537",
    archivePrefix = "arXiv",
    primaryClass = "hep-th",
    doi = "10.1007/JHEP12(2015)071",
    journal = "JHEP",
    volume = "12",
    pages = "071",
    year = "2015"
}

@article{Alvarez-Gaume:2016vff,
    author = "Alvarez-Gaume, Luis and Loukas, Orestis and Orlando, Domenico and Reffert, Susanne",
    title = "{Compensating strong coupling with large charge}",
    eprint = "1610.04495",
    archivePrefix = "arXiv",
    primaryClass = "hep-th",
    reportNumber = "CERN-TH-2016-221",
    doi = "10.1007/JHEP04(2017)059",
    journal = "JHEP",
    volume = "04",
    pages = "059",
    year = "2017"
}

@article{Komargodski:2012ek,
    author = "Komargodski, Zohar and Zhiboedov, Alexander",
    title = "{Convexity and Liberation at Large Spin}",
    eprint = "1212.4103",
    archivePrefix = "arXiv",
    primaryClass = "hep-th",
    doi = "10.1007/JHEP11(2013)140",
    journal = "JHEP",
    volume = "11",
    pages = "140",
    year = "2013"
}

@article{Fitzpatrick:2012yx,
    author = "Fitzpatrick, A. Liam and Kaplan, Jared and Poland, David and Simmons-Duffin, David",
    title = "{The Analytic Bootstrap and AdS Superhorizon Locality}",
    eprint = "1212.3616",
    archivePrefix = "arXiv",
    primaryClass = "hep-th",
    doi = "10.1007/JHEP12(2013)004",
    journal = "JHEP",
    volume = "12",
    pages = "004",
    year = "2013"
}

@article{Cuomo:2023mxg,
    author = "Cuomo, Gabriel and Lopes, J. M. Viana Parente and Matos, Jos\'e and Oliveira, J\'ulio and Penedones, Joao",
    title = "{Numerical tests of the large charge expansion}",
    eprint = "2305.00499",
    archivePrefix = "arXiv",
    primaryClass = "hep-lat",
    doi = "10.1007/JHEP05(2024)161",
    journal = "JHEP",
    volume = "05",
    pages = "161",
    year = "2024"
}

@article{Badel:2022fya,
    author = "Badel, Gil and Monin, Alexander and Rattazzi, Riccardo",
    title = "{Identifying Large Charge operators}",
    eprint = "2207.08919",
    archivePrefix = "arXiv",
    primaryClass = "hep-th",
    doi = "10.1007/JHEP02(2023)119",
    journal = "JHEP",
    volume = "02",
    pages = "119",
    year = "2023"
}

@article{Dondi:2024vua,
    author = "Dondi, Nicola Andrea and Sberveglieri, Giacomo",
    title = "{NLO in the large charge sector of the critical $O(N)$ model at large $N$}",
    eprint = "2409.06781",
    archivePrefix = "arXiv",
    primaryClass = "hep-th",
    month = "9",
    year = "2024"
}

@article{Alday:2007mf,
    author = "Alday, Luis F. and Maldacena, Juan Martin",
    title = "{Comments on operators with large spin}",
    eprint = "0708.0672",
    archivePrefix = "arXiv",
    primaryClass = "hep-th",
    doi = "10.1088/1126-6708/2007/11/019",
    journal = "JHEP",
    volume = "11",
    pages = "019",
    year = "2007"
}

@article{Cuomo:2019ejv,
    author = "Cuomo, Gabriel",
    title = "{Superfluids, vortices and spinning charged operators in 4d CFT}",
    eprint = "1906.07283",
    archivePrefix = "arXiv",
    primaryClass = "hep-th",
    doi = "10.1007/JHEP02(2020)119",
    journal = "JHEP",
    volume = "02",
    pages = "119",
    year = "2020"
}

@article{Kravec:2019djc,
    author = "Kravec, S. M. and Pal, Sridip",
    title = "{The Spinful Large Charge Sector of Non-Relativistic CFTs: From Phonons to Vortex Crystals}",
    eprint = "1904.05462",
    archivePrefix = "arXiv",
    primaryClass = "hep-th",
    doi = "10.1007/JHEP05(2019)194",
    journal = "JHEP",
    volume = "05",
    pages = "194",
    year = "2019"
}

@article{Cuomo:2023vvd,
    author = "Cuomo, Gabriel and Komargodski, Zohar and Zhong, Siwei",
    title = "{Chiral modes of giant superfluid vortices}",
    eprint = "2312.06095",
    archivePrefix = "arXiv",
    primaryClass = "cond-mat.quant-gas",
    doi = "10.1103/PhysRevB.110.144514",
    journal = "Phys. Rev. B",
    volume = "110",
    number = "14",
    pages = "144514",
    year = "2024"
}

@article{Moroz:2018noc,
    author = "Moroz, Sergej and Hoyos, Carlos and Benzoni, Claudio and Son, Dam Thanh",
    title = "{Effective field theory of a vortex lattice in a bosonic superfluid}",
    eprint = "1803.10934",
    archivePrefix = "arXiv",
    primaryClass = "cond-mat.quant-gas",
    doi = "10.21468/SciPostPhys.5.4.039",
    journal = "SciPost Phys.",
    volume = "5",
    number = "4",
    pages = "039",
    year = "2018"
}

@article{Cooper_2008,
   title={Rapidly rotating atomic gases},
   volume={57},
   ISSN={1460-6976},
   url={http://dx.doi.org/10.1080/00018730802564122},
   DOI={10.1080/00018730802564122},
   number={6},
   journal={Advances in Physics},
   publisher={Informa UK Limited},
   author={Cooper, N.R.},
   year={2008},
   month=nov, pages={539–616} }

@phdthesis{Pellizzani:2024gfa,
    author = "Pellizzani, Vito",
    title = "{The large-charge expansion in nonrelativistic conformal field theories}",
    eprint = "2502.17224",
    archivePrefix = "arXiv",
    primaryClass = "hep-th",
    doi = "10.48549/5455",
    school = "Bern U.",
    year = "2024"
}

@article{Choi:2025tql,
    author = "Choi, Jaehyeok and Lee, Eunwoo",
    title = "{Large charge operators at large spin from relativistically rotating vortices}",
    eprint = "2501.07198",
    archivePrefix = "arXiv",
    primaryClass = "hep-th",
    month = "1",
    year = "2025"
}

@article{Watanabe:2013iia,
    author = "Watanabe, Haruki and Murayama, Hitoshi",
    title = "{Redundancies in Nambu-Goldstone Bosons}",
    eprint = "1302.4800",
    archivePrefix = "arXiv",
    primaryClass = "cond-mat.other",
    reportNumber = "IPMU13-0046",
    doi = "10.1103/PhysRevLett.110.181601",
    journal = "Phys. Rev. Lett.",
    volume = "110",
    number = "18",
    pages = "181601",
    year = "2013"
}

@article{Fetter:2009zz,
    author = "Fetter, Alexander L.",
    title = "{Rotating trapped Bose-Einstein condensates}",
    doi = "10.1103/RevModPhys.81.647",
    journal = "Rev. Mod. Phys.",
    volume = "81",
    pages = "647--691",
    year = "2009"
}

@article{Son:2005rv,
    author = "Son, D. T. and Wingate, M.",
    title = "{General coordinate invariance and conformal invariance in nonrelativistic physics: Unitary Fermi gas}",
    eprint = "cond-mat/0509786",
    archivePrefix = "arXiv",
    reportNumber = "INT-PUB-05-23",
    doi = "10.1016/j.aop.2005.11.001",
    journal = "Annals Phys.",
    volume = "321",
    pages = "197--224",
    year = "2006"
}

@article{Blau:2009gd,
    author = "Blau, Matthias and Hartong, Jelle and Rollier, Blaise",
    title = "{Geometry of Schrodinger Space-Times, Global Coordinates, and Harmonic Trapping}",
    eprint = "0904.3304",
    archivePrefix = "arXiv",
    primaryClass = "hep-th",
    doi = "10.1088/1126-6708/2009/07/027",
    journal = "JHEP",
    volume = "07",
    pages = "027",
    year = "2009"
}

@article{Son:2008ye,
    author = "Son, D. T.",
    title = "{Toward an AdS/cold atoms correspondence: A Geometric realization of the Schrodinger symmetry}",
    eprint = "0804.3972",
    archivePrefix = "arXiv",
    primaryClass = "hep-th",
    reportNumber = "INT-PUB-08-08",
    doi = "10.1103/PhysRevD.78.046003",
    journal = "Phys. Rev. D",
    volume = "78",
    pages = "046003",
    year = "2008"
}

@article{Balasubramanian:2008dm,
    author = "Balasubramanian, Koushik and McGreevy, John",
    title = "{Gravity duals for non-relativistic CFTs}",
    eprint = "0804.4053",
    archivePrefix = "arXiv",
    primaryClass = "hep-th",
    doi = "10.1103/PhysRevLett.101.061601",
    journal = "Phys. Rev. Lett.",
    volume = "101",
    pages = "061601",
    year = "2008"
}

@article{Herzog:2008wg,
    author = "Herzog, Christopher P. and Rangamani, Mukund and Ross, Simon F.",
    title = "{Heating up Galilean holography}",
    eprint = "0807.1099",
    archivePrefix = "arXiv",
    primaryClass = "hep-th",
    reportNumber = "PUPT-2274, DCPT-08-39",
    doi = "10.1088/1126-6708/2008/11/080",
    journal = "JHEP",
    volume = "11",
    pages = "080",
    year = "2008"
}

@article{Maldacena:2008wh,
    author = "Maldacena, Juan and Martelli, Dario and Tachikawa, Yuji",
    title = "{Comments on string theory backgrounds with non-relativistic conformal symmetry}",
    eprint = "0807.1100",
    archivePrefix = "arXiv",
    primaryClass = "hep-th",
    doi = "10.1088/1126-6708/2008/10/072",
    journal = "JHEP",
    volume = "10",
    pages = "072",
    year = "2008"
}

@article{Adams:2008wt,
    author = "Adams, Allan and Balasubramanian, Koushik and McGreevy, John",
    title = "{Hot Spacetimes for Cold Atoms}",
    eprint = "0807.1111",
    archivePrefix = "arXiv",
    primaryClass = "hep-th",
    reportNumber = "MIT-CTP-3962",
    doi = "10.1088/1126-6708/2008/11/059",
    journal = "JHEP",
    volume = "11",
    pages = "059",
    year = "2008"
}

@article{Rangamani:2008gi,
    author = "Rangamani, Mukund and Ross, Simon F. and Son, D. T. and Thompson, Ethan G.",
    title = "{Conformal non-relativistic hydrodynamics from gravity}",
    eprint = "0811.2049",
    archivePrefix = "arXiv",
    primaryClass = "hep-th",
    reportNumber = "DCPT-08-61, INT-PUB-08-48",
    doi = "10.1088/1126-6708/2009/01/075",
    journal = "JHEP",
    volume = "01",
    pages = "075",
    year = "2009"
}

@article{Anand:2025mfh,
    author = "Anand, Harsh and Benjamin, Nathan and Kumar, Vipul and Minwalla, Shiraz and Mukherjee, Jyotirmoy and Pal, Sridip and Rahaman, Asikur",
    title = "{Semi-universality of CFT$_d$ entropy at large spin}",
    eprint = "2512.00158",
    archivePrefix = "arXiv",
    primaryClass = "hep-th",
    month = "11",
    year = "2025"
}

@article{Nishida:2007pj,
    author = "Nishida, Yusuke and Son, Dam T.",
    title = "{Nonrelativistic conformal field theories}",
    eprint = "0706.3746",
    archivePrefix = "arXiv",
    primaryClass = "hep-th",
    reportNumber = "INT-PUB-07-16",
    doi = "10.1103/PhysRevD.76.086004",
    journal = "Phys. Rev. D",
    volume = "76",
    pages = "086004",
    year = "2007"
}

@article{Geracie:2014nka,
    author = "Geracie, Michael and Son, Dam Thanh and Wu, Chaolun and Wu, Shao-Feng",
    title = "{Spacetime Symmetries of the Quantum Hall Effect}",
    eprint = "1407.1252",
    archivePrefix = "arXiv",
    primaryClass = "cond-mat.mes-hall",
    reportNumber = "EFI-14-21",
    doi = "10.1103/PhysRevD.91.045030",
    journal = "Phys. Rev. D",
    volume = "91",
    pages = "045030",
    year = "2015"
}

@article{Kravec:2018qnu,
    author = "Kravec, S. M. and Pal, Sridip",
    title = "{Nonrelativistic Conformal Field Theories in the Large Charge Sector}",
    eprint = "1809.08188",
    archivePrefix = "arXiv",
    primaryClass = "hep-th",
    doi = "10.1007/JHEP02(2019)008",
    journal = "JHEP",
    volume = "02",
    pages = "008",
    year = "2019"
}

@article{Son:2005ak,
    author = "Son, D. T.",
    title = "{Effective Lagrangian and topological interactions in supersolids}",
    eprint = "cond-mat/0501658",
    archivePrefix = "arXiv",
    reportNumber = "INT-PUB-05-06",
    doi = "10.1103/PhysRevLett.94.175301",
    journal = "Phys. Rev. Lett.",
    volume = "94",
    pages = "175301",
    year = "2005"
}

@article{Tkachenko1965,
  author  = {V. K. Tkachenko},
  title   = {On Vortex Lattices in a Rotating Superfluid},
  journal = {Zhurnal Eksperimental'noi i Teoreticheskoi Fiziki},
  volume  = {49},
  pages   = {1875--1882},
  year    = {1965},
  note    = {Sov. Phys. JETP 22, 1282 (1966)}
}

@article{Bajaj:2024utv,
    author = "Bajaj, Kabir and Kumar, Vipul and Minwalla, Shiraz and Mukherjee, Jyotirmoy and Rahaman, Asikur",
    title = "{Grey Galaxies in $AdS_5$}",
    eprint = "2412.06904",
    archivePrefix = "arXiv",
    primaryClass = "hep-th",
    reportNumber = "TIFR/TH/24-25",
    month = "12",
    year = "2024"
}

@article{Banerjee:2012iz,
    author = "Banerjee, Nabamita and Bhattacharya, Jyotirmoy and Bhattacharyya, Sayantani and Jain, Sachin and Minwalla, Shiraz and Sharma, Tarun",
    title = "{Constraints on Fluid Dynamics from Equilibrium Partition Functions}",
    eprint = "1203.3544",
    archivePrefix = "arXiv",
    primaryClass = "hep-th",
    reportNumber = "TFR-TH-12-05, IPMU12-0037",
    doi = "10.1007/JHEP09(2012)046",
    journal = "JHEP",
    volume = "09",
    pages = "046",
    year = "2012"
}

@article{Jensen:2012jh,
    author = "Jensen, Kristan and Kaminski, Matthias and Kovtun, Pavel and Meyer, Rene and Ritz, Adam and Yarom, Amos",
    title = "{Towards hydrodynamics without an entropy current}",
    eprint = "1203.3556",
    archivePrefix = "arXiv",
    primaryClass = "hep-th",
    reportNumber = "CCTP-2012-03",
    doi = "10.1103/PhysRevLett.109.101601",
    journal = "Phys. Rev. Lett.",
    volume = "109",
    pages = "101601",
    year = "2012"
}

@article{Cordova:2017dhq,
    author = "Cordova, Clay and Diab, Kenan",
    title = "{Universal Bounds on Operator Dimensions from the Average Null Energy Condition}",
    eprint = "1712.01089",
    archivePrefix = "arXiv",
    primaryClass = "hep-th",
    doi = "10.1007/JHEP02(2018)131",
    journal = "JHEP",
    volume = "02",
    pages = "131",
    year = "2018"
}

@article{Lee:2025qim,
    author = "Lee, Eunwoo",
    title = "{Extremal AdS Black Holes as Fluids: A Matrix Large-Charge EFT Approach}",
    eprint = "2507.21240",
    archivePrefix = "arXiv",
    primaryClass = "hep-th",
    month = "7",
    year = "2025"
}
\end{document}